# Three-dimensional vibrations of multilayered hollow spheres submerged in a complex fluid

B. Wu[1,2,3,4], Y. Gan[4], E. Carrera[3,*], W. Q. Chen[1,2,4,5,*]

[1] Key Laboratory of Soft Machines and Smart Devices of Zhejiang Province,
Zhejiang University, Hangzhou 310027, P.R. China
[2] State Key Lab of CAD & CG, Zhejiang University, Hangzhou 310058, China
[3] Department of Mechanical and Aerospace Engineering,
Politecnico di Torino, Torino 10129, Italy
[4] Department of Engineering Mechanics, Zhejiang University, Hangzhou 310027, China
[5] Soft Matter Research Center, Zhejiang University, Hangzhou 310027, P.R. China

Fluid-structure interaction is fundamental to the characteristics of the induced flows due to the motion of structures in fluids and also is crucial to the performance of submerged structures. This paper presents a three-dimensional analytical study of the intrinsic free vibration of an elastic multilayered hollow sphere interacting with an exterior non-Newtonian fluid medium. The fluid is assumed to be characterized by a compressible linear viscoelastic model accounting for both the shear and compressional relaxation processes. For small-amplitude vibrations, the equations governing the viscoelastic fluid can be linearized, which are then solved by introducing appropriate potential functions. The solid is assumed to exhibit a particular material anisotropy, i.e. spherical isotropy, which includes material isotropy as a special case. The equations governing the anisotropic solid are solved in spherical coordinates using the state-space formalism, which finally establishes two separate transfer relations correlating the state vectors at the innermost surface with those at the outermost surface of the multilayered hollow sphere. By imposing the continuity conditions at the fluid-solid interface, two separate analytical characteristic equations are derived, which characterize two independent classes of vibration. Numerical examples are finally conducted to validate the theoretical derivation as well as to investigate the effects of various factors, including fluid viscosity and compressibility, fluid viscoelasticity, solid anisotropy and surface effect, as well as solid intrinsic damping, on the vibration characteristics of the submerged hollow sphere. Particularly, our theoretically predicted vibration frequencies and quality factors of gold nanospheres with intrinsic damping immersed in water agree exceptionally well with the available experimentally measured results. The reported analytical solution is truly and fully three-dimensional, covering from the purely radial breathing mode to torsional mode to any general spheroidal mode as well as being applicable to various simpler situations, and hence can be a broad-spectrum benchmark in the study of fluid-structure interaction.

## 1. Introduction

Acoustic vibrations of elastic bodies in fluids have long been an interesting topic that is also of

---

significant practical importance in many applications, such as designing underwater acoustic wave devices, characterizing mechanical properties of small-scale materials, exploiting mechanical and biological sensing mechanism, probing fluid environment, regulating flow pattern, and even developing methods to destroy viruses. The fluid-structure interaction (FSI) is known to play a key role in affecting the motional behavior of either the surrounding fluid medium or the submerged elastic body. There already have been a number of comprehensive monographs (e.g. Junger & Feit 1986; Howe 1998; Païdoussis 2003) and excellent review papers (e.g. Dowell & Hall 2001; Hou, Wang & Layton 2012) on this topic, to which the reader may be referred. Some recent interesting works on vortex-induced vibrations of an elastically mounted sphere are also mentioned here; these include Behara, Borazjani & Sotiropoulos (2011), Sareen *et al*. (2018), Rajamuni, Thompson & Hourigan (2018).

A major part of the existing literature on FSI focuses on the interactions between homogeneous isotropic elastic bodies and Newtonian fluids. In other words, the classical Navier-Stokes equations (in terms of velocities) are adopted to capture the flow behavior of the fluid medium (e.g. Batchelor 1967), while the classical Navier equations (in terms of displacements) are employed for the solid medium (e.g. Timoshenko & Goodier 1973). Since the pioneering works of Faran (1951) and Junger (1952a, 1952b), numerous experimental, numerical, and analytical studies have been carried out to better understand the interaction mechanisms and the resulting effects. Among them, analytical analysis is of particular research interest, especially for solids with simple geometry, such as cylinder, sphere, and plate. For example, for a thin elastic spherical shell submerged in a compressible fluid, Ding and Chen (1998) verified in a mathematically strict sense for the first time that any frequency of non-torsional free vibrations shall be complex because of the energy dissipation associated with the disturbance in fluid that propagates away from the vibrating shell. Complex frequencies are a common aspect for the free vibration of a structure submerged or embedded in another infinite medium due to the same energy dissipation mechanism except for a few particular cases (e.g. the torsional vibration of a sphere in an inviscid fluid).

The scope of FSI study has been broadened in two ways regarding the physical models. The first one generalizes the basic assumptions made on the material properties of structures or solids, which are allowed to be inhomogeneous and/or anisotropic. For example, Ding and Chen (1996) considered the coupled vibrations of an elastic hollow sphere which possesses spherical isotropy, a material symmetry with a total of five independent material constants. Chen, Wang & Ding (1999) further assumed that the sphere is radially functionally graded so that the material constants vary continuously along the radial direction. In addition, the solid may feature a multi-field coupling effect such as the piezoelectric materials and piezomagnetic materials (e.g. Zhu *et al.* 2013; Jiang, Zhu & Chen 2019). Nonlinear elastic models have also been used to describe the solid medium in the case that its nonlinear behavior should be considered (e.g. Gao, Howard & Ponte Castañeda 2011; Khaderi & Onck 2012; Nasouri, Khot & Elfring 2017). The logic behind this kind of expansion of the FSI study is the increasing applications of advanced functional materials in modern technologies.

The other way is to adopt more complex fluid models, i.e. from Newtonian fluids to non-Newtonian fluids. There are also definite driving forces to do so. One is clearly associated with the situations where non-Newtonian fluids are present (Lauga & Powers 2009; Janela, Moura & Sequeira 2010; Liao 2003; Li & Ardekani 2015; Boyko, Bercovici & Gat 2017; Dey, Modarres-Sadeghi & Rothstein 2017). Another one deals with acoustic vibrations of nanoparticles (e.g. Chakraborty *et al.* 2013) or viruses (e.g. Babincová, Sourivong & Babinec 2000; Saviot, Netting & Murray 2007). Pelton *et al.* (2013) carried out an elaborate experiment to measure the dynamic responses of a bipyramidal gold nanoparticle immersed in glycerol-water mixtures, and they found that the quality factor of acoustic oscillation of the nanoparticle first decreases and then increases with the mass fraction of glycerol. This phenomenon is opposed to the prediction based on the Newtonian fluid model, but agrees quite well with that from a non-Newtonian fluid model. Since then, a few studies have been reported on the acoustic vibrations of nanoparticles interacting with non-Newtonian fluids (e.g. Yu *et al.* 2015; Chakraborty & Sader 2015; Chakraborty *et al.* 2018). It shall be noted that new phenomena that have not been observed before may manifest themselves in experiments of acoustic vibrations of nanoparticles or viruses at extremely high frequencies in the range of GHz to THz. The fluid viscoelastic properties are only one aspect; another one is dewetting of the nanoparticle that was recently revealed by Hsueh, Gordon & Rottler (2018) using molecular dynamics simulations.

From the analytical aspect, Galstyan, Pak & Stone (2015) derived an exact characteristic equation for the purely radial vibration or breathing mode of a spherical nanoparticle in a linear Maxwell fluid. They showed that the quality factor of the nanosphere varies in a similar way to that of a bipyramidal gold nanoparticle observed by Pelton *et al.* (2013). Chakraborty & Sader (2015) extended their discussion to other models of non-Newtonian fluids, and illustrated the importance of both the shear and compressional relaxation processes in the non-Newtonian fluid, but still focusing on the breathing mode for the spherical nanoparticle. However, different modes of nanoparticles may be excited under different external stimuli (Liu *et al.* 2009; Crut *et al.* 2015), and hence it is quite interesting and also important to develop an analytical approach to the general three-dimensional (3D) vibration. Historically, the 3D analytical solution of free vibration of an isotropic elastic sphere immersed in a viscoelastic material was obtained much earlier by Saviot, Netting & Murray (2007). But there are no 3D analytical solutions reported for inhomogeneous and anisotropic nanoparticles interacting with a non-Newtonian fluid. It is the purpose of this study to present such a general treatment by noticing three important aspects: 1) the silver or gold nanoparticles are cubic in nature with three-independent material constants, which cannot be described by an isotropic elastic model (Crut *et al.* 2009); 2) at nanoscales, the surface effect usually becomes important (Cammarata 1997; Wu, Chen & Zhang 2018), and also surface adhesion layer could be effectively utilized to tune the dynamic responses of a nanoparticle (Chang *et al.* 2015); and 3) experimentally measured quality factors of different nanoparticles are considerably lower than the theoretical predictions (Ruijgrok *et al.* 2012; Chakraborty *et al.* 2013), which suggests that additional damping mechanisms intrinsic to the solid particles also should be examined.

This paper develops a comprehensive theoretical analysis of the 3D coupled vibration of a multilayered anisotropic hollow sphere submerged in an infinite non-Newtonian fluid characterized by a compressional linear viscoelastic model proposed by Yong (2014) and Chakraborty & Sader (2015), as illustrated in figure 1. Full fluid-structure interaction is taken into consideration by imposing the continuity conditions at the fluid-solid interface. Numerical examples are finally conducted to investigate the effects of different factors, including fluid viscosity and compressibility, fluid viscoelasticity, solid anisotropy, solid surface effect, and solid intrinsic damping, on the vibration characteristics of the FSI system. The study can be envisioned as a generalization of the existing works mentioned above. The obtained 3D analytical solution covers a few simpler and degenerated cases. For example, the general 3D vibration naturally embraces the torsional modes and the spheroidal modes (e.g. breathing, dipolar and quadrupolar modes); the multilayered structure can be made homogeneous by assuming uniform material properties; the hollow configuration can be degenerated to the solid one by taking a very small inner radius; a very thin outermost layer can be used to model the surface effect or the surface adhesion/capping layer; an isotropic material is a particular case of the material with spherical isotropy; and of course, the compressible linear viscoelastic fluid can reduce to the linear Maxwell fluid and Newtonian fluid, either compressible or incompressible, viscous or non-viscous. Thus, the presented 3D analytical solution can play a versatile role in clarifying approximate or numerical approaches as well as interpreting experimental results in the FSI study.

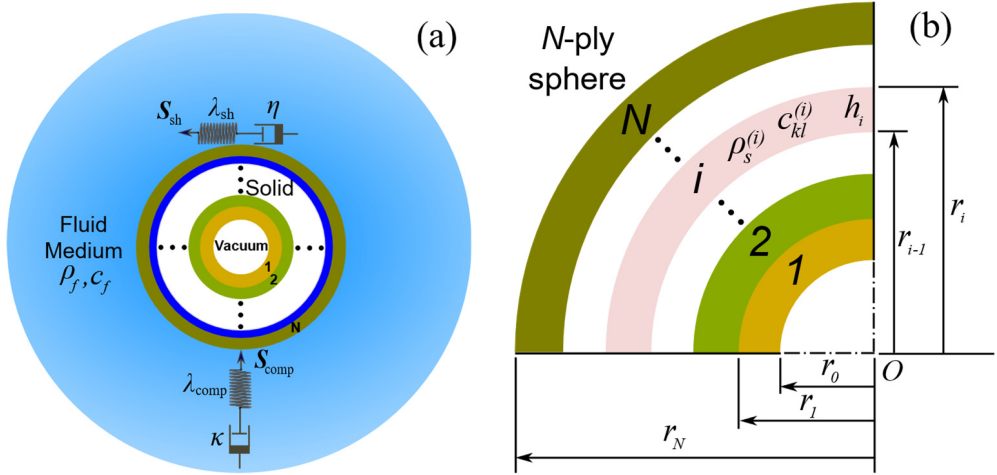

FIGURE 1. (a) Schematic diagram of a multilayered hollow sphere submerged in an infinite non-Newtonian fluid accounting for both shear and compressional relaxation effects; (b) Geometric schematic of an anisotropic *N*-ply hollow sphere.

## 2. Harmonic analysis of the exterior non-Newtonian fluid

Pelton *et al.* (2013) analytically and experimentally studied the high-frequency longitudinal vibration of a bipyramidal gold nanoparticle based on an incompressible linear Maxwell model which neglects the effects of both liquid compressibility and compressional relaxation spectrum due to the shear-dominated nature of the flow. For the purely radial vibration or breathing mode of a spherical gold nanoparticle, Galstyan, Pak & Stone (2015) used a compressible linear Maxwell model to investigate the effects of both liquid compressibility and viscoelasticity in simple liquids. However, the compressible linear Maxwell model only considers the shear relaxation effect and overlooks the compressional relaxation process which may play a significant role in the dynamic response of a pressure wave in a non-Newtonian fluid. As a mathematical and physical modification to the linear Maxwell model, Yong (2014) and Chakraborty & Sader (2015) proposed a compressible linear viscoelastic model accounting for both shear and compressional relaxation effects, which is characterized by a spring-dashpot connected in series for each of the shear and compressional relaxation processes, as displayed in figure 1(a). The compressible linear viscoelastic model will be referred to as *compressional non-Newtonian model* afterwards for simplicity, which can recover a purely Newtonian result in the low frequency limit and reproduce a purely elastic response in the high frequency limit (Chakraborty & Sader 2015). In this paper, we assume that the fluid is non-Newtonian, and can be described by the compressional non-Newtonian model. Furthermore, the linearized governing equations of the compressible viscoelastic fluid are solved by introducing appropriate potential functions in the case of small-amplitude vibrations.

Since small-amplitude vibrations are considered, all time-dependent perturbations of the density, pressure and velocity are assumed small and all convective terms can be neglected. Therefore, the linearized continuity equation for a compressible fluid is given by

$$\frac{\partial \rho'}{\partial t} + \rho_f \nabla \cdot \mathbf{v} = 0, \tag{1}$$

where $\nabla$ is the usual del operator, $\rho_f$ is the fluid density at equilibrium, and $\rho'$ denotes the density perturbation with $\rho' \ll \rho_f$; the fluid velocity field $\mathbf{v}$ satisfies the following linearized equations of motion:

$$\rho_f \frac{\partial \mathbf{v}}{\partial t} = \nabla \cdot \boldsymbol{\sigma}^f, \tag{2}$$

where the Cauchy stress tensor in fluid can be written as $\boldsymbol{\sigma}^f = -p\mathbf{I} + \mathbf{S}$ with $p$ denoting the thermodynamic pressure, $\mathbf{I}$ being the identity tensor, and $\mathbf{S}$ representing the extra stress tensor (Lai, Krempl & Rubin 2009). In general, the extra stress tensor can be expressed as $\mathbf{S} = \mathbf{S}_{sh} + \mathbf{S}_{comp}$, where the deviatoric stress tensor $\mathbf{S}_{sh}$ has a vanishing trace (Galstyan, Pak & Stone 2015), while the spherical stress tensor $\mathbf{S}_{comp}$ is an isotropic tensor (Yong 2014). Therefore, these two tensors are orthogonal, i.e., $\mathbf{S}_{sh} : \mathbf{S}_{comp} = 0$, which means that $\mathbf{S}_{sh}$ and $\mathbf{S}_{comp}$ represent the shear and compressional contributions to the extra stress tensor, respectively.

In order to describe both the shear and compressional relaxation processes of a compressible linear viscoelastic fluid, $\mathbf{S}_{sh}$ and $\mathbf{S}_{comp}$ are related to the rate-of-strain tensor $\mathbf{D} = [\nabla \mathbf{v} + (\nabla \mathbf{v})^T]/2$ by

$$\mathbf{S}_{sh} + \lambda_{sh} \frac{\partial \mathbf{S}_{sh}}{\partial t} = 2\eta \left( \mathbf{D} - \frac{\text{tr}\mathbf{D}}{3} \mathbf{I} \right), \quad \mathbf{S}_{comp} + \lambda_{comp} \frac{\partial \mathbf{S}_{comp}}{\partial t} = \kappa (\text{tr}\mathbf{D}) \mathbf{I}, \tag{3}$$

where $\text{tr}\mathbf{D} = \nabla \cdot \mathbf{v}$, $\lambda_{sh}$ and $\lambda_{comp}$ specify the shear and compressional relaxation times of the fluid, respectively, and $\eta$ and $\kappa$ are the fluid shear and bulk viscosities, respectively. The constitutive relation (3) is the compressional non-Newtonian model proposed by Yong (2014) and Chakraborty & Sader (2015). Neglecting the compressional relaxation effect of the fluid (i.e., let $\lambda_{comp} = 0$), the constitutive model (3) reduces to the classical linear Maxwell model. If one further ignores the shear relaxation process (i.e., let $\lambda_{sh} = 0$), the standard compressible Newtonian constitutive model (Lai, Krempl & Rubin 2009) can be obtained, i.e., $\mathbf{S}_{sh} = 2\eta[\mathbf{D} - (\mathrm{tr}\mathbf{D})\mathbf{I}/3]$ and $\mathbf{S}_{comp} = \kappa(\mathrm{tr}\mathbf{D})\mathbf{I}$.

Focusing on oscillatory motions, we seek harmonic solutions with the explicit time dependence $\mathrm{e}^{-\mathrm{i}\omega t}$ for all time-dependent variables such as the fluid velocity, pressure, and stresses. Here "i" is the imaginary unit and $\omega$ is the circular frequency of vibration. Therefore, the constitutive relation (3), with such a time dependence, can be rewritten as

$$\mathbf{S}_{sh} = 2\eta'\left(\mathbf{D} - \frac{\mathrm{tr}\mathbf{D}}{3}\mathbf{I}\right), \quad \mathbf{S}_{comp} = \kappa'(\mathrm{tr}\mathbf{D})\mathbf{I}, \tag{4}$$

where $\eta' = \eta/(1-\mathrm{i}\omega\lambda_{sh})$ and $\kappa' = \kappa/(1-\mathrm{i}\omega\lambda_{comp})$. Comparing Eq. (4) with the standard Newtonian constitutive relation, one can observe that the harmonic solutions to the compressional non-Newtonian model (3) can be rigorously obtained from those of the compressible Newtonian model by simply making the substitutions $\eta \to \eta'$ and $\kappa \to \kappa'$.

The linearized equation-of-state relating the fluid density perturbation $\rho'$ to the thermodynamic pressure $p$ can be written as $p = c_f^2 \rho'$, where $c_f$ is the speed of sound in the fluid at thermodynamic equilibrium. Substitution of Eq. (4) into Eq. (2) yields the following generalized Navier-Stokes equation for the compressional non-Newtonian model (3) as

$$\rho_f \frac{\partial \mathbf{v}}{\partial t} = -\nabla p + \eta'\nabla^2 \mathbf{v} + \left(\kappa' + \frac{\eta'}{3}\right)\nabla(\nabla \cdot \mathbf{v}). \tag{5}$$

where $\nabla^2$ is the 3D Laplacian.

It is well known that the fluid velocity vector, satisfying Eq. (5), can be expressed by a dilatational scalar potential $\Phi$ and an equivoluminal vector potential $\mathbf{\Psi}$, such that

$$\mathbf{v} = \nabla \times \mathbf{\Psi} + \nabla\Phi, \tag{6}$$

where $\Phi$ and $\mathbf{\Psi}$ satisfy the following equations (Guz 1980)

$$\left[\frac{1}{c_f^2}\frac{\partial^2}{\partial t^2} - \left(1 + \frac{\kappa' + 4\eta'/3}{c_f^2 \rho_f}\frac{\partial}{\partial t}\right)\nabla^2\right]\Phi = 0, \quad \left(\frac{\partial}{\partial t} - \frac{\eta'}{\rho_f}\nabla^2\right)\mathbf{\Psi} = \mathbf{0}. \tag{7}$$

In the spherical coordinate system $(r,\theta,\varphi)$ with the unit basis vectors $\mathbf{e}_r$, $\mathbf{e}_\theta$, and $\mathbf{e}_\varphi$, the solution to Eq. (7)$_2$ can be represented as (Guz 1980)

$$\Psi = r\Psi_1 \mathbf{e}_r + \nabla \times (r\Psi_2 \mathbf{e}_r), \tag{8}$$

where the two scalar potentials $\Psi_1$ and $\Psi_2$ are determined from the following equations:

$$\left(\frac{\partial}{\partial t} - \frac{\eta'}{\rho_f}\nabla^2\right)\Psi_j = 0, \quad j = 1, 2. \tag{9}$$

In the spherical coordinate system, we have $\nabla^2 = \partial^2/\partial r^2 + (2/r)\partial/\partial r + (1/r^2)\nabla_1^2$, where the two-dimensional Laplacian is expressed as $\nabla_1^2 = \partial^2/\partial\theta^2 + (\cot\theta)\partial/\partial\theta + (1/\sin^2\theta)\partial^2/\partial\varphi^2$.

The fluid density perturbation $\rho'$ and the thermodynamic pressure $p$ are thus obtained from the following expressions:

$$\rho' = \frac{\rho_f}{c_f^2}\left(\frac{\kappa' + 4\eta'/3}{\rho_f}\nabla^2 - \frac{\partial}{\partial t}\right)\Phi, \quad p = \rho_f\left(\frac{\kappa' + 4\eta'/3}{\rho_f}\nabla^2 - \frac{\partial}{\partial t}\right)\Phi. \tag{10}$$

For the harmonic motion, the Cauchy stress tensor in fluid for the compressional non-Newtonian model (3), in view of Eq. (4), is

$$\boldsymbol{\sigma}^f = -p\mathbf{I} + 2\eta'\left(\mathbf{D} - \frac{\mathrm{tr}\mathbf{D}}{3}\mathbf{I}\right) + \kappa'(\mathrm{tr}\mathbf{D})\mathbf{I}. \tag{11}$$

Therefore, all time-dependent variables in Eqs. (6), (10) and (11) for the compressional non-Newtonian model (3) can be expressed in terms of the three scalar potentials $\Phi$, $\Psi_1$ and $\Psi_2$ which are determined from Eqs. (7)$_1$ and (9).

For the problem under consideration, the fluid scalar potentials can be expanded in series in the following form:

$$\Phi = e^{-i\omega t}\sum_{n=0}^{\infty}\sum_{m=-n}^{n}\phi_n(r)X_n^m(\theta,\varphi), \quad \Psi_j = e^{-i\omega t}\sum_{n=0}^{\infty}\sum_{m=-n}^{n}\psi_{jn}(r)X_n^m(\theta,\varphi), \tag{12}$$

where $X_n^m(\theta,\varphi) = P_n^m(\cos\theta)e^{im\varphi}$ are the spherical harmonic functions, $P_n^m(\cos\theta)$ are the associated Legendre polynomials, and the integers $n$ and $m$ specify the angular and azimuthal dependences of the vibration modes, respectively. Note that it is not necessary to indicate the integer $m$ in the subscript of the unknown functions $\phi_n$ and $\psi_{jn}$ in Eq. (12) since it will not appear in the resulting ordinary differential equations about these functions, as to be shown in the following derivations. Substituting Eq. (12) into Eqs. (7)$_1$ and (9), one can obtain the governing equations of $\phi_n$ and $\psi_{jn}$ as

$$\left(\frac{1}{r^2}\frac{\mathrm{d}}{\mathrm{d}r}r^2\frac{\mathrm{d}}{\mathrm{d}r}-\frac{n(n+1)}{r^2}+\alpha_1^2\right)\phi_n=0, \quad \left(\frac{1}{r^2}\frac{\mathrm{d}}{\mathrm{d}r}r^2\frac{\mathrm{d}}{\mathrm{d}r}-\frac{n(n+1)}{r^2}+\alpha_2^2\right)\psi_{jn}=0, \tag{13}$$

where the identity $\nabla_1^2 X_n^m(\theta,\varphi)=-n(n+1)X_n^m(\theta,\varphi)$ has been used and

$$\alpha_1^2=\omega^2 c_f^{-2}\left(1-\mathrm{i}\omega c_f^{-2}\frac{\kappa'+4\eta'/3}{\rho_f}\right)^{-1}, \quad \alpha_2^2=\mathrm{i}\omega\frac{\rho_f}{\eta'}. \tag{14}$$

For the coupled vibration of submerged closed spheres or spherical shells, all generated waves are going outwards, which requires that the solutions to the three scalar potentials in Eq. (13) take the following forms:

$$\phi_n(r)=A_n h_n^{(1)}(\alpha_1 r), \quad \psi_{jn}(r)=B_{jn}h_n^{(1)}(\alpha_2 r), \tag{15}$$

where $h_n^{(1)}$ is the spherical Hankel function of the first kind of order $n$, and $A_n$ and $B_{jn}$ are arbitrary constants. Consequently, the components of velocity vector and Cauchy stress tensor can be derived in terms of the three scalar potentials $\Phi$, $\Psi_1$ and $\Psi_2$. Inserting Eq. (12) into Eqs. (8) and (6) yields the components of the velocity vector $\mathbf{v}$ as

$$\begin{aligned}
v_r &= \mathrm{e}^{-\mathrm{i}\omega t}\sum_{n=0}^{\infty}\sum_{m=-n}^{n}\left[\frac{\mathrm{d}\phi_n}{\mathrm{d}r}+\frac{n(n+1)}{r}\psi_{2n}\right]X_n^m(\theta,\varphi), \\
v_\theta &= \mathrm{e}^{-\mathrm{i}\omega t}\sum_{n=0}^{\infty}\sum_{m=-n}^{n}\left[\left(\frac{\phi_n}{r}+\frac{1}{r}\frac{\mathrm{d}}{\mathrm{d}r}(r\psi_{2n})\right)\frac{\partial X_n^m}{\partial \theta}+\frac{\psi_{1n}}{\sin\theta}\frac{\partial X_n^m}{\partial \varphi}\right], \\
v_\varphi &= \mathrm{e}^{-\mathrm{i}\omega t}\sum_{n=0}^{\infty}\sum_{m=-n}^{n}\left[\left(\frac{\phi_n}{r}+\frac{1}{r}\frac{\mathrm{d}}{\mathrm{d}r}(r\psi_{2n})\right)\frac{1}{\sin\theta}\frac{\partial X_n^m}{\partial \varphi}-\psi_{1n}\frac{\partial X_n^m}{\partial \theta}\right].
\end{aligned} \tag{16}$$

Substituting Eq. (12)$_1$ into Eq. (10)$_2$, and inserting the obtained results and Eq. (16) into Eq. (11) give the components of the Cauchy stress tensor $\boldsymbol{\sigma}^f$ on a spherical surface (with the normal unit vector $\mathbf{e}_r$) as

$$\begin{aligned}
\frac{\sigma_{rr}^f}{2\eta'} &= \mathrm{e}^{-\mathrm{i}\omega t}\sum_{n=0}^{\infty}\sum_{m=-n}^{n}\left[\left(\alpha_1^2-\mathrm{i}\omega\frac{\rho_f}{2\eta'}+\frac{\mathrm{d}^2}{\mathrm{d}r^2}\right)\phi_n+n(n+1)\frac{\mathrm{d}}{\mathrm{d}r}\left(\frac{\psi_{2n}}{r}\right)\right]X_n^m(\theta,\varphi), \\
\frac{\sigma_{r\theta}^f}{2\eta'} &= \mathrm{e}^{-\mathrm{i}\omega t}\sum_{n=0}^{\infty}\sum_{m=-n}^{n}\left\{\frac{1}{\sin\theta}\frac{\partial X_n^m}{\partial \varphi}\frac{r}{2}\frac{\mathrm{d}}{\mathrm{d}r}\left(\frac{\psi_{1n}}{r}\right)+\frac{\partial X_n^m}{\partial \theta}\left[\frac{\mathrm{d}}{\mathrm{d}r}\left(\frac{\phi_n}{r}\right)-\left(\frac{1}{r}\frac{\mathrm{d}}{\mathrm{d}r}-\frac{n(n+1)-1}{r^2}+\frac{\alpha_2^2}{2}\right)\psi_{2n}\right]\right\}, \\
\frac{\sigma_{r\varphi}^f}{2\eta'} &= \mathrm{e}^{-\mathrm{i}\omega t}\sum_{n=0}^{\infty}\sum_{m=-n}^{n}\left\{\frac{1}{\sin\theta}\frac{\partial X_n^m}{\partial \varphi}\left[\frac{\mathrm{d}}{\mathrm{d}r}\left(\frac{\phi_n}{r}\right)-\left(\frac{1}{r}\frac{\mathrm{d}}{\mathrm{d}r}-\frac{n(n+1)-1}{r^2}+\frac{\alpha_2^2}{2}\right)\psi_{2n}\right]-\frac{\partial X_n^m}{\partial \theta}\frac{r}{2}\frac{\mathrm{d}}{\mathrm{d}r}\left(\frac{\psi_{1n}}{r}\right)\right\},
\end{aligned} \tag{17}$$

where Eq. (13)$_1$ has been used in the derivation of $\sigma_{rr}^f$.

With Eqs. (15)-(17), the continuity conditions at the fluid-solid interface will provide direct interactions between the compressional non-Newtonian fluid and the closed sphere or spherical shell which will be discussed in Secs. 3 and 4.

## 3. Vibration of the multilayered anisotropic hollow sphere

Lamb (1881) presented the first 3D solution for the free vibration of an isotropic elastic sphere in vacuum. Saviot, Netting & Murray (2007) extended Lamb's analysis to the case when the exterior medium is an infinite viscous compressible Newtonian fluid. On the other hand, Cohen, Shah & Ramakrishnan (1972) made a generalization of Lamb's analysis by assuming that the solid is spherically isotropic. Ding and Chen (1996) further analytically studied the effects of exterior non-viscous compressible fluids on the free vibrations of spherically isotropic hollow spheres. For the free vibration of laminated spheres, Chen and Ding (2001) developed a more efficient approach based on the state-space formalism along with the Taylor's expansion. To the authors' best knowledge, no 3D analytical solutions are available in the open literature for free vibrations of inhomogeneous and anisotropic spheres interacting with a non-Newtonian fluid.

For a spherically isotropic elastic hollow sphere, the spherical coordinates $(r,\theta,\varphi)$ are used with the origin $O$ located at the center of the spherical isotropy and the linear constitutive relations are given by

$$\begin{aligned}
\Sigma_{\theta\theta} &= r\sigma_{\theta\theta} = c_{11}E_{\theta\theta} + c_{12}E_{\varphi\varphi} + c_{13}E_{rr}, \quad \Sigma_{\varphi\varphi} = r\sigma_{\varphi\varphi} = c_{12}E_{\theta\theta} + c_{11}E_{\varphi\varphi} + c_{13}E_{rr}, \\
\Sigma_{rr} &= r\sigma_{rr} = c_{13}E_{\theta\theta} + c_{13}E_{\varphi\varphi} + c_{33}E_{rr}, \quad \Sigma_{r\theta} = r\sigma_{r\theta} = 2c_{44}E_{r\theta}, \\
\Sigma_{r\varphi} &= r\sigma_{r\varphi} = 2c_{44}E_{r\varphi}, \quad \Sigma_{\theta\varphi} = r\sigma_{\theta\varphi} = 2c_{66}E_{\theta\varphi},
\end{aligned} \quad (18)$$

where $\sigma_{ij}$ is the stress tensor, $\Sigma_{ij}$ is the generalized stress tensor, $c_{ij}$ are the elastic constants, and the relation $c_{66} = (c_{11} - c_{12})/2$ holds for spherical isotropy. The generalized strain tensor $E_{ij}$ in Eq. (18) can be written as follows:

$$\begin{aligned}
E_{rr} &= re_{rr} = \nabla_2 u_r, \quad E_{\theta\theta} = re_{\theta\theta} = \frac{\partial u_\theta}{\partial \theta} + u_r, \quad E_{\varphi\varphi} = re_{\varphi\varphi} = \frac{1}{\sin\theta}\frac{\partial u_\varphi}{\partial \varphi} + u_r + u_\theta \cot\theta, \\
2E_{r\theta} &= 2re_{r\theta} = \frac{\partial u_r}{\partial \theta} + \nabla_2 u_\theta - u_\theta, \quad 2E_{r\varphi} = 2re_{r\varphi} = \frac{1}{\sin\theta}\frac{\partial u_r}{\partial \varphi} + \nabla_2 u_\varphi - u_\varphi, \\
2E_{\theta\varphi} &= 2re_{\theta\varphi} = \frac{1}{\sin\theta}\frac{\partial u_\theta}{\partial \varphi} + \frac{\partial u_\varphi}{\partial \theta} - u_\varphi \cot\theta,
\end{aligned} \quad (19)$$

where $\nabla_2 = r\partial/\partial r$, $u_i \ (i = r,\theta,\varphi)$ are the three displacement components, and $e_{ij}$ is the strain tensor. In spherical coordinates, the differential equations of motion without body forces can be easily transformed into the following form in terms of the generalized stress tensor $\Sigma_{ij}$:

$$\nabla_2 \Sigma_{r\theta} + \frac{1}{\sin\theta}\frac{\partial \Sigma_{\theta\varphi}}{\partial \varphi} + \frac{\partial \Sigma_{\theta\theta}}{\partial \theta} + 2\Sigma_{r\theta} + (\Sigma_{\theta\theta} - \Sigma_{\varphi\varphi})\cot\theta = \rho_s r^2 \frac{\partial^2 u_\theta}{\partial t^2},$$

$$\nabla_2 \Sigma_{r\varphi} + \frac{1}{\sin\theta}\frac{\partial \Sigma_{\varphi\varphi}}{\partial \varphi} + \frac{\partial \Sigma_{\theta\varphi}}{\partial \theta} + 2\Sigma_{r\varphi} + 2\Sigma_{\theta\varphi}\cot\theta = \rho_s r^2 \frac{\partial^2 u_\varphi}{\partial t^2}, \qquad (20)$$

$$\nabla_2 \Sigma_{rr} + \frac{1}{\sin\theta}\frac{\partial \Sigma_{r\varphi}}{\partial \varphi} + \frac{\partial \Sigma_{r\theta}}{\partial \theta} + \Sigma_{rr} - \Sigma_{\theta\theta} - \Sigma_{\varphi\varphi} + \Sigma_{r\theta}\cot\theta = \rho_s r^2 \frac{\partial^2 u_r}{\partial t^2},$$

where $\rho_s$ is the mass density of the solid.

In this work, the state-space formalism along with the approximate laminate technique will be exploited to solve the 3D governing equations (18)-(20). By directly choosing the state variables as $(u_r, u_\theta, u_\varphi, \sigma_{rr}, \sigma_{r\theta}, \sigma_{r\varphi})$, it is not difficult to establish the corresponding state equation (Wu, Chen & Zhang 2018). However, it has been shown by Chen & Ding (2001) that, by introducing proper potential functions, not only can the basic governing equations be decoupled with the order reduced, the subsequent solving procedure also becomes simpler. In fact, one may introduce the following three displacement functions $F$, $G$, $w$ and two stress functions $\Sigma_1$, $\Sigma_2$ as

$$u_\theta = -\frac{1}{\sin\theta}\frac{\partial F}{\partial \varphi} - \frac{\partial G}{\partial \theta}, \quad u_\varphi = \frac{\partial F}{\partial \theta} - \frac{1}{\sin\theta}\frac{\partial G}{\partial \varphi}, \quad u_r = w,$$

$$\Sigma_{r\theta} = -\frac{1}{\sin\theta}\frac{\partial \Sigma_1}{\partial \varphi} - \frac{\partial \Sigma_2}{\partial \theta}, \quad \Sigma_{r\varphi} = \frac{\partial \Sigma_1}{\partial \theta} - \frac{1}{\sin\theta}\frac{\partial \Sigma_2}{\partial \varphi}. \qquad (21)$$

Substituting Eq. (21) into Eqs. (18)-(20), through some lengthy mathematical manipulations, one can transform the original 3D equations into two separated state equations, both with variable coefficients. The reader is referred to the Supplemental Material (SM)-I for the derivations of the state equations for the general non-axisymmetric free vibration of a closed *N*-ply hollow sphere shown in figure 1(b).

Since it is intractable to solve the state equations with variable coefficients directly, the approximate laminate technique (Chen & Ding 2002; Ding, Chen & Zhang 2006; Wu *et al.* 2017) can be employed to derive the approximate analytical solutions and then obtain the relation between the state vectors at the inner and outer surfaces of the *N*-ply hollow sphere (see SM-I for more information of the approximate laminate technique). As a result, two separate transfer relations can be established as

$$\mathbf{T}_{kn}^{\text{in}} = \mathbf{H}_{kn}\mathbf{T}_{kn}^{\text{ou}}, \quad (k \in \{1,2\}; \ n \geq 1), \qquad (22)$$

where $\mathbf{T}_{kn}^{\text{in}}$ and $\mathbf{T}_{kn}^{\text{ou}}$ are the dimensionless state vectors at the innermost and outermost surfaces of the $N$-ply hollow sphere, respectively, and $\mathbf{H}_{kn}$ are the global transfer matrices of second-order ($k=1$) and fourth-order ($k=2$). The dimensionless state vectors at arbitrary point in the $N$-ply hollow sphere are $\mathbf{T}_{1n} = [T_{1n1}, T_{1n2}]^{\mathrm{T}}$ and $\mathbf{T}_{2n} = [T_{2n1}, T_{2n2}, T_{2n3}, T_{2n4}]^{\mathrm{T}}$ with their components (i.e. state variables) defined as

$$T_{1n1} = \Sigma_{1n}/(r_N c_{44}^{(1)}), \quad T_{1n2} = F_n/r_N, \quad T_{2n1} = \Sigma_{rn}/(r_N c_{44}^{(1)}),$$
$$T_{2n2} = \Sigma_{2n}/(r_N c_{44}^{(1)}), \quad T_{2n3} = G_n/r_N, \quad T_{2n4} = w_n/r_N, \tag{23}$$

where $r_N$ denotes the outermost radius of the $N$-ply hollow sphere, the superscript '(1)' of $c_{44}^{(1)}$ indicates the material constant of the first layer, and $F_n$, $G_n$, $w_n$, $\Sigma_{1n}$, $\Sigma_{2n}$ and $\Sigma_{rn}$ are six unknown functions depending on the radial coordinate $r$. It is obvious from Eq. (23) that the two unknown functions $F_n$ and $\Sigma_{1n}$ are uncoupled from the other four unknown functions $G_n$, $w_n$, $\Sigma_{2n}$ and $\Sigma_{rn}$.

For the breathing mode $n=0$ corresponding to the purely radial vibration, Eq. (22) degenerates to the following transfer relation:

$$\{T_{201}^{\mathrm{in}}, T_{204}^{\mathrm{in}}\}^{\mathrm{T}} = \mathbf{H}_{20} \{T_{201}^{\mathrm{ou}}, T_{204}^{\mathrm{ou}}\}^{\mathrm{T}}, \quad (n=0), \tag{24}$$

where $\mathbf{H}_{20}$ is a second-order global transfer matrix.

By incorporating the fluid-solid interface conditions, one can readily obtain the frequency equations of 3D free vibrations of the multilayered hollow sphere interacting with an exterior fluid medium, which will be discussed in detail in Sec. 4.

## 4. Fluid-solid interface conditions and frequency equations

Before coupling the vibrational motion of the $N$-ply hollow sphere to the flow of the surrounding fluid, we first utilize the similar separation formulae to Eq. (21) to express the velocities and radial stresses of the fluid as

$$v_\theta = -\frac{1}{\sin\theta}\frac{\partial F^f}{\partial \varphi} - \frac{\partial G^f}{\partial \theta}, \quad v_\varphi = -\frac{1}{\sin\theta}\frac{\partial G^f}{\partial \varphi} + \frac{\partial F^f}{\partial \theta}, \quad v_r = w^f,$$
$$r\sigma_{r\theta}^f = -\frac{1}{\sin\theta}\frac{\partial \Sigma_1^f}{\partial \varphi} - \frac{\partial \Sigma_2^f}{\partial \theta}, \quad r\sigma_{r\varphi}^f = \frac{\partial \Sigma_1^f}{\partial \theta} - \frac{1}{\sin\theta}\frac{\partial \Sigma_2^f}{\partial \varphi}, \quad r\sigma_{rr}^f = \Sigma_{rr}^f, \tag{25}$$

where the superscript "$f$" represents the quantities belonging to the fluid. For harmonic vibration solutions, comparing Eq. (25) with Eqs. (16) and (17) as well as utilizing the orthogonality of the associated Legendre polynomials with respect to the weight $\sin\theta$ (Ding, Chen & Zhang 2006), one can obtain the following relations between the (velocity and stress) potential functions $F_n^f$, $G_n^f$, $w_n^f$, $\Sigma_{rn}^f$, $\Sigma_{1n}^f$ and $\Sigma_{2n}^f$ for the fluid case and the three scalar potentials $\phi_n$ and $\psi_{jn}$ of the fluid medium:

$$F_n^f = -\psi_{1n}, \quad -\frac{\Sigma_{1n}^f}{r} = 2\eta'\frac{r}{2}\frac{d}{dr}\left(\frac{\psi_{1n}}{r}\right), \tag{26}$$

$$w_n^f = \frac{d\phi_n}{dr} + \frac{n(n+1)}{r}\psi_{2n}, \quad G_n^f = -\left[\frac{\phi_n}{r} + \frac{1}{r}\frac{d}{dr}(r\psi_{2n})\right],$$

$$\frac{\Sigma_{rn}^f}{r} = 2\eta'\left[\left(\alpha_1^2 - i\omega\frac{\rho_f}{2\eta'} + \frac{d^2}{dr^2}\right)\phi_n + n(n+1)\frac{d}{dr}\left(\frac{\psi_{2n}}{r}\right)\right], \tag{27}$$

$$-\frac{\Sigma_{2n}^f}{r} = 2\eta'\left[\frac{d}{dr}\left(\frac{\phi_n}{r}\right) - \left(\frac{1}{r}\frac{d}{dr} - \frac{n(n+1)-1}{r^2} + \frac{\alpha_2^2}{2}\right)\psi_{2n}\right].$$

We now turn to the boundary conditions and fluid-solid interface conditions of the submerged $N$-ply hollow sphere. At the inner spherical surface $r = r_0$, the free boundary conditions require that the three surface stresses vanish, i.e., $\sigma_{rr} = \sigma_{r\theta} = \sigma_{r\varphi} = 0$, which can be expressed in terms of the state variables as

$$\Sigma_{1n} = \Sigma_{rn} = \Sigma_{2n} = 0, \quad r = r_0. \tag{28}$$

At the outer spherical surface $r = r_N$ (i.e., the fluid-solid interface), the velocities and three stress components along the radial direction should be continuous, i.e.,

$$\dot{u}_r = v_r, \quad \dot{u}_\theta = v_\theta, \quad \dot{u}_\varphi = v_\varphi, \quad \sigma_{rr} = \sigma_{rr}^f, \quad \sigma_{r\theta} = \sigma_{r\theta}^f, \quad \sigma_{r\varphi} = \sigma_{r\varphi}^f, \quad r = r_N, \tag{29}$$

which provides direct interaction between the sphere and the fluid. Using Eqs. (13)$_1$, (15), (21), (25)-(27) and the derivative formula of the spherical Hankel function of the first kind $d[h_n^{(1)}(x)]/dx = nh_n^{(1)}(x)/x - h_{n+1}^{(1)}(x)$, the fluid-solid interface conditions (29) can be rewritten as

$$-i\omega F_n = -B_{1n}h_n^{(1)}(\alpha_2 r_N), \quad \Sigma_{1n} = -\eta' B_{1n}\left[(n-1)h_n^{(1)}(\alpha_2 r_N) - \alpha_2 r_N h_{n+1}^{(1)}(\alpha_2 r_N)\right], \tag{30}$$

$$\begin{aligned}
-\mathrm{i}\omega w_n &= A_n\left[\frac{n}{r_N}h_n^{(1)}(\alpha_1 r_N) - \alpha_1 h_{n+1}^{(1)}(\alpha_1 r_N)\right] + B_{2n}\frac{n(n+1)}{r_N}h_n^{(1)}(\alpha_2 r_N), \\
-\mathrm{i}\omega G_n &= -\left\{A_n\frac{1}{r_N}h_n^{(1)}(\alpha_1 r_N) + B_{2n}\left[\frac{n+1}{r_N}h_n^{(1)}(\alpha_2 r_N) - \alpha_2 h_{n+1}^{(1)}(\alpha_2 r_N)\right]\right\}, \\
\Sigma_{rn} &= 2\eta'\left\{\left[\left(\frac{n(n-1)}{r_N} - \mathrm{i}\omega\frac{\rho_f}{2\eta'}r_N\right)h_n^{(1)}(\alpha_1 r_N) + 2\alpha_1 h_{n+1}^{(1)}(\alpha_1 r_N)\right]A_n \right. \\
&\quad \left. + B_{2n}n(n+1)\left[\frac{n-1}{r_N}h_n^{(1)}(\alpha_2 r_N) - \alpha_2 h_{n+1}^{(1)}(\alpha_2 r_N)\right]\right\}, \\
\Sigma_{2n} &= -2\eta'\left\{A_n\left[\frac{n-1}{r_N}h_n^{(1)}(\alpha_1 r_N) - \alpha_1 h_{n+1}^{(1)}(\alpha_1 r_N)\right] \right. \\
&\quad \left. - B_{2n}\left[\left(\frac{1-n^2}{r_N} + \frac{r_N\alpha_2^2}{2}\right)h_n^{(1)}(\alpha_2 r_N) - \alpha_2 h_{n+1}^{(1)}(\alpha_2 r_N)\right]\right\}.
\end{aligned} \qquad (31)$$

It is clear from Eqs. (28) and (30)-(31) that, for the coupled vibration of the submerged multilayered hollow sphere, both the free boundary conditions and the fluid-solid interface conditions can be separated into two categories: one is associated with three functions $F_n$, $\Sigma_{1n}$, and $\psi_{1n}$ (see Eqs. (28)$_1$ and (30)), while the other is related to the other six functions $w_n$, $G_n$, $\Sigma_{rn}$, $\Sigma_{2n}$, $\phi_n$ and $\psi_{2n}$ (see Eqs. (28)$_{2,3}$ and (31)).

Using the results obtained above, the 3D free vibrations of a multilayered hollow sphere interacting with an exterior compressional non-Newtonian fluid may be divided into two independent classes, just like the case without the FSI (Lamb 1881; Chen & Ding 2001; Ding, Chen & Zhang 2006). Specifically, the first class $(n \geq 1)$ [defined by Eqs. (13)$_2$ for $j=1$, (22) for $k=1$, and (30)] corresponds to torsional (or toroidal) modes and belongs to an equi-volumetric motion (i.e., $\mathrm{div}\,\mathbf{u}=0$) of the multilayered hollow sphere without the radial displacement, while for the second class $(n \geq 0)$ [governed by Eqs. (13)$_1$, (13)$_2$ for $j=2$, (22) for $k=1$, (24) and (31)], the mechanical displacements generally possess both transverse and radial components except for the breathing mode, for which only the radial component is nonzero. The second class of vibration corresponds to the spheroidal modes including, for instances, the breathing ($n=0$), dipolar ($n=1$) and quadrupolar ($n=2$) modes, which may be excited under different external stimuli (Liu *et al.* 2009; Crut *et al.* 2015).

Making using of the dimensionless state variables defined in Eq. (23), one can nondimensionalize the inner and outer surface boundary conditions (28) and (30)-(31) as

$$T_{1n1}^{\text{in}} = T_{2n1}^{\text{in}} = T_{2n2}^{\text{in}} = 0, \tag{32}$$

and

$$T_{1n1}^{\text{ou}} = \bar{B}_{1n} R_1, \quad T_{1n2}^{\text{ou}} = \bar{B}_{1n} R_2, \tag{33}$$

$$\begin{aligned} T_{2n1}^{\text{ou}} &= \bar{A}_n Q_{11} + \bar{B}_{2n} Q_{12}, \quad T_{2n2}^{\text{ou}} = \bar{A}_n Q_{21} + \bar{B}_{2n} Q_{22}, \\ T_{2n3}^{\text{ou}} &= \bar{A}_n Q_{31} + \bar{B}_{2n} Q_{32}, \quad T_{2n4}^{\text{ou}} = \bar{A}_n Q_{41} + \bar{B}_{2n} Q_{42}, \end{aligned} \tag{34}$$

where the dimensionless undetermined coefficients $\bar{A}_n$, $\bar{B}_{1n}$ and $\bar{B}_{2n}$ are defined as

$$\bar{A}_n = \frac{A_n h_n^{(1)}(\tilde{\alpha}_1)}{r_N \sqrt{c_{44}^{(1)}/\rho_s^{(1)}}}, \quad \bar{B}_{2n} = \frac{B_{2n} h_n^{(1)}(\tilde{\alpha}_2)}{r_N \sqrt{c_{44}^{(1)}/\rho_s^{(1)}}}, \quad \bar{B}_{1n} = \frac{B_{1n} h_n^{(1)}(\tilde{\alpha}_2)}{\sqrt{c_{44}^{(1)}/\rho_s^{(1)}}}, \tag{35}$$

and

$$\begin{aligned}
R_1 &= -\frac{\tilde{\rho}}{\tilde{r}'\tilde{v}}\left[n - 1 - \tilde{\alpha}_2 g_n(\tilde{\alpha}_2)\right], \quad R_2 = -\frac{\text{i}}{\Omega}, \\
Q_{11} &= \frac{2\tilde{\rho}}{\tilde{r}'\tilde{v}}\left[n(n-1) - \text{i}\Omega\frac{\tilde{r}'\tilde{v}}{2} + 2\tilde{\alpha}_1 g_n(\tilde{\alpha}_1)\right], \quad Q_{12} = \frac{2\tilde{\rho}}{\tilde{r}'\tilde{v}} n(n+1)\left[n - 1 - \tilde{\alpha}_2 g_n(\tilde{\alpha}_2)\right], \\
Q_{21} &= -\frac{2\tilde{\rho}}{\tilde{r}'\tilde{v}}\left[n - 1 - \tilde{\alpha}_1 g_n(\tilde{\alpha}_1)\right], \quad Q_{22} = \frac{2\tilde{\rho}}{\tilde{r}'\tilde{v}}\left[1 - n^2 + \frac{\tilde{\alpha}_2^2}{2} - \tilde{\alpha}_2 g_n(\tilde{\alpha}_2)\right], \\
Q_{31} &= -\frac{\text{i}}{\Omega}, \quad Q_{32} = -\frac{\text{i}}{\Omega}\left[n + 1 - \tilde{\alpha}_2 g_n(\tilde{\alpha}_2)\right], \\
Q_{41} &= \frac{\text{i}}{\Omega}\left[n - \tilde{\alpha}_1 g_n(\tilde{\alpha}_1)\right], \quad Q_{42} = \frac{\text{i}}{\Omega} n(n+1), \quad g_n(x) = \frac{h_{n+1}^{(1)}(x)}{h_n^{(1)}(x)},
\end{aligned} \tag{36}$$

in which we have introduced the following dimensionless parameters:

$$\tilde{\alpha}_1 = \alpha_1 r_N = \Omega \tilde{v}\left(1 - \frac{\gamma' + 4/3}{\tilde{r}'}\text{i}\Omega \tilde{v}\right)^{-1/2}, \quad \tilde{\alpha}_2 = \alpha_2 r_N = \sqrt{\text{i}\Omega \tilde{r}'\tilde{v}}, \quad \tilde{\rho} = \frac{\rho_f}{\rho_s^{(1)}}, \tag{37}$$

$$\tilde{v} = \frac{\sqrt{c_{44}^{(1)}/\rho_s^{(1)}}}{c_f}, \quad \gamma' = \frac{\kappa'}{\eta'} = \gamma \frac{1 - \text{i}\Omega \tilde{\lambda}_{\text{sh}}}{1 - \text{i}\Omega \tilde{\lambda}_{\text{comp}}}, \quad \tilde{r}' = \frac{\rho_f c_f r_N}{\eta'} = \tilde{r}(1 - \text{i}\Omega \tilde{\lambda}_{\text{sh}}),$$

where

$$\Omega = r_N \omega \sqrt{\frac{\rho_s^{(1)}}{c_{44}^{(1)}}}, \quad \gamma = \frac{\kappa}{\eta}, \quad \tilde{r} = \frac{\rho_f c_f r_N}{\eta}, \quad \tilde{\lambda}_{\text{sh}} = \frac{\lambda_{\text{sh}}\sqrt{c_{44}^{(1)}/\rho_s^{(1)}}}{r_N}, \quad \tilde{\lambda}_{\text{comp}} = \frac{\lambda_{\text{comp}}\sqrt{c_{44}^{(1)}/\rho_s^{(1)}}}{r_N}. \tag{38}$$

It should be underlined that the dimensionless parameters defined in Eqs. (37)-(38) have clear physical meanings: 1) $\Omega$ indicates the normalized circular frequency, which is complex due to the presence of damping mechanisms such as fluid viscosity and compressibility, while $\tilde{\alpha}_1$ and $\tilde{\alpha}_2$ are the normalized wave numbers, both along the radial direction; 2) the density ratio $\tilde{\rho}$ represents the relative strength of fluid-to-solid inertia (i.e., the added fluid mass effect), whereas $\gamma$ gives the bulk-to-shear viscosity ratio; 3) $\tilde{v}$ specifies the ratio of elastic wave velocity in solid to the speed of sound in fluid, which stands for the effect of fluid compressibility (i.e., the fluid is incompressible when $c_f$ is taken to be infinity); 4) the effects of shear and compressional viscoelasticity are controlled by $\tilde{\lambda}_{sh}$ and $\tilde{\lambda}_{comp}$, and thus fluid elasticity may be neglected when they are both very small; 5) the effect of shear viscosity is governed by the *quasi Reynolds number* $\tilde{r}$ (in terms of the sound speed $c_f$ rather than the fluid velocity $|\mathbf{v}|$), large values of which stands for a negligible effect of fluid viscosity. In addition, the dimensionless parameters with a prime (i.e., $\eta'$, $\kappa'$, $\gamma'$, and $\tilde{r}'$) are adopted for the compressional non-Newtonian model; they are merely the counterparts of the Newtonian fluid parameters $\eta$, $\kappa$, $\gamma$, and $\tilde{r}$.

Substituting Eqs. (32)-(34) into Eqs. (22) and (24), we can obtain three sets of linear homogeneous algebraic equations for the undetermined constants $\bar{A}_n$, $\bar{B}_{1n}$ and $\bar{B}_{2n}$. As a necessary condition for nontrivial solutions, the determinant of coefficients of each set of algebraic equations should vanish. This results in the characteristic frequency equations of two independent classes of vibrations, respectively,

$$H_{1n11}R_1 + H_{1n12}R_2 = 0, \quad (n \geq 1) \tag{39}$$

for the first class (torsional modes), and

$$H_{2n11}Q_{11} + H_{2n12}Q_{41} = 0, \quad (n = 0), \tag{40}$$

$$\begin{vmatrix} U_{2n11} & U_{2n12} \\ U_{2n21} & U_{2n22} \end{vmatrix} = 0, \quad (n \geq 1) \tag{41}$$

for the second class (spheroidal modes), where

$$U_{2n11} = \sum_{k=1}^{4} H_{2n1k}Q_{k1}, \; U_{2n12} = \sum_{k=1}^{4} H_{2n1k}Q_{k2}, \; U_{2n21} = \sum_{k=1}^{4} H_{2n2k}Q_{k1}, \; U_{2n22} = \sum_{k=1}^{4} S_{2n2k}Q_{k2}. \tag{42}$$

In Eqs. (39)-(42), $H_{knij}$ represents the element on the *i*th row and *j*th column of the coefficient matrix $\mathbf{H}_{kn}$. It is emphasized here that, the azimuthal integer *m* representing the non-axisymmetric vibration modes does not appear in the resulting characteristic frequency equations (39)-(42). The underlying physical mechanism is apparent since any non-axisymmetric vibration mode can be obtained by the superposition of different axisymmetric ones with identical natural frequencies with respect to differently oriented spherical coordinates (Chen & Ding 2001).

Once the vibration frequency is found from Eqs. (39)-(42), the unknown state variables at the inner and outer surfaces can be determined from the corresponding algebraic equations. The state variables at any interior point in the multilayered hollow sphere can be calculated accordingly. The three induced variables $\Sigma_{\theta\varphi}$, $\Sigma_{\theta\theta}$, and $\Sigma_{\phi\phi}$ of the multilayered hollow sphere can be also determined from the state variables. The reader is referred to SM-I for more details.

It should be pointed out that, if the compressional relaxation process is discarded (i.e., let $\lambda_{\text{comp}} = 0$), the obtained 3D analytical solutions for a compressional non-Newtonian fluid reduce to those for a linear Maxwell fluid. When the shear relaxation effect is further neglected (i.e., let $\lambda_{\text{sh}} = 0$), the results of a viscous compressible Newtonian fluid can be recovered. For some simpler degenerated cases including non-viscous compressible Newtonian fluids, incompressible linear viscoelastic fluids and non-viscous incompressible fluids, their characteristic frequency equations of two independent classes of vibrations of the FSI system have been derived by exploiting the asymptotic analysis (see SM-II for more details).

## 5. Numerical examples and discussions

We now conduct numerical calculations for a (multilayered) solid sphere submerged in a complex fluid in order to quantitatively investigate the effects of different factors, including fluid viscosity and compressibility, fluid viscoelasticity, solid anisotropy and surface effect, as well as solid intrinsic damping on the vibration characteristics of the FSI system.

Due to the presence of different damping mechanisms, the circular frequency in general will be complex, and can be written as $\omega = \omega_r - \mathrm{i}\omega_i$ with $\omega_r$ and $-\omega_i$ being the real and imaginary parts of the frequency. The corresponding normalized counterparts are denoted as $\Omega_r$ and $-\Omega_i$ according to the relation $\Omega = r_N \omega \sqrt{\rho_s^{(1)} / c_{44}^{(1)}}$. In the following, by "vibration frequency" we mean the real part of frequency, while by "damping component" we mean the imaginary part. In addition, the quality factor is defined as $Q = \sqrt{\omega_r^2 + \omega_i^2} / (2\omega_i)$ following the definition adopted in Galstyan, Pak & Stone (2015). Furthermore, it should be mentioned that since the frequency equations (39)-(42) are derived for 3D motions, there exist an infinite number of eigen-frequencies (i.e., the number of radial nodes or the harmonic order in Lamb's theory varies) for each angular mode number $n$.

In this paper, the obtained frequency equations (39)-(42) are solved by the *Muller's method* (Muller 1956). The Muller's method is an iterative algorithm, which generalizes the secant method but uses quadratic interpolation among three points instead of linear interpolation between two points. Thus it is also referred to as the *parabolic method*. Solving for the zeros of the quadratic equation enables us to find the complex pairs of roots.

### 5.1. *Comparisons with the available theoretical and experimental results*

TABLE 1. Comparison of the quality factor $Q$ and vibration frequency $f = \omega_r / 2\pi$ of the fundamental (or first-order) breathing mode with available theoretical predictions for isotropic gold nanospheres (40nm and 10nm radii) vibrating in non-viscous compressible, viscous compressible, linear Maxwell, and compressional non-Newtonian fluids. The fluid media include the pure water ($\chi = 0$) and glycerol-water mixture ($\chi = 0.56$), and their material constants are referred to SM-III for details. The exact theoretical predictions are made by Galstyan, Pak & Stone (2015).

| Fluid medium | Nonvis. Comp. | | Vis. Comp. | | Linear Max. | | Comp. non-New. |
|---|---|---|---|---|---|---|---|
| | SSM | Exact* | SSM | Exact | SSM | Exact | SSM |
| Water (40nm) | 53.7 (37.9 GHz) | 53.7 (37.9 GHz) | 52.6 (37.8 GHz) | 52.6 (37.8 GHz) | 52.2 (37.8 GHz) | 52.2 (37.8 GHz) | 50.8 (37.8 GHz) |
| Mixture (40nm) | 40.6 (37.9 GHz) | 40.6 (37.9 GHz) | 33.4 (37.6 GHz) | 33.4 (37.6 GHz) | 34.5 (37.7 GHz) | 34.5 (37.7 GHz) | 30.6 (37.8 GHz) |
| Water (10nm) | 53.7 (151.7GHz) | 53.7 (151.7GHz) | 43.5 (150.7GHz) | 43.5 (150.7GHz) | 43.8 (151.0GHz) | 43.8 (151.0GHz) | 38.1 (151.2GHz) |
| Mixture (10nm) | 40.6 (151.5GHz) | 40.6 (151.5GHz) | 18.6 (149.2GHz) | 18.6 (149.2GHz) | 33.9 (150.3GHz) | 33.9 (150.3GHz) | 27.3 (151.5GHz) |

The analysis based on the state-space formalism combined with the approximate laminate technique (hereafter abbreviated as *state-space method or SSM*) along with the Muller's method will be verified in this subsection in terms of the accuracy by comparisons with the available theoretical and experimental results. Employing the proposed method, we first calculate the vibration frequency $f = \omega_r / 2\pi$ and the quality factor $Q$ of the fundamental (or first-order) breathing mode ($n = 0$) of isotropic gold nanospheres of two different radii (40nm and 10nm) vibrating in two glycerol-water mixtures. The numerical results are given in table 1 for non-viscous compressible, viscous compressible, linear Maxwell, and compressional non-Newtonian fluid models. The theoretical predictions based on the conventional displacement method in Galstyan, Pak & Stone (2015) are also presented in table 1 for comparison and validation purposes. As described in Sec. 2, the linear Maxwell model without the compressional relaxation process was utilized by Galstyan, Pak & Stone (2015) to study the breathing mode. Thus, the results predicted by the compressional non-Newtonian model are also shown in table 1 to highlight the compressional relaxation effect. It can be seen that the calculated results based on SSM agree exceptionally well with those by the exact displacement method. In addition, the compressional relaxation process will have a significant influence on the vibration quality factor, especially for the nanosphere with a smaller radius and the fluid medium with a larger viscoelasticity.

TABLE 2. Comparison of the vibration frequency $f = \omega_r / 2\pi$ for various experimentally excited and observed modes for different isotropic nanospheres vibrating in different surrounding media. The experimental measurements concerning a 40nm gold nanoparticles in water and a 50nm free influenza A virions are conducted by Ruijgrok *et al.* (2012) and Liu *et al.* (2009). The experimental data from the literature are displayed in square brackets.

| Different nanospheres | Breathing (1st order) | Breathing (2nd order) | Breathing (3rd order) | Quadrupolar (1st order) | Dipolar (1st order) | Dipolar (2st order) |
|---|---|---|---|---|---|---|
| Gold in water (40nm) | 37.8 GHz [38.8 GHz] | 80 GHz [84 GHz] | 121 GHz [126 GHz] | 12.7 GHz [12.5 GHz] | — | — |
| Influenza A virions (50nm) | — | — | — | — | 14 GHz [12 GHz] | 28 GHz [26 GHz] |

Combining ultrafast pump-probe spectroscopy with optical trapping, Ruijgrok *et al.* (2012) conducted an elaborate experiment to measure the dynamic responses of a single gold nanoparticle (40nm radius) immersed in water and detected the fundamental and higher-order breathing modes and the first-order quadrupolar mode. Through the dipolar coupling to confined acoustic modes, the microwave resonant absorption (MRA) of virions was confirmed by Liu *et al.* (2009) and the absorption peaks of the first two dipolar modes was observed for influenza A virions (with 50nm radius, effective longitudinal wave velocity 2450m/s, and Poisson's ratio 0.33). Hence, the theoretically predicted vibration frequency based on the compressional non-Newtonian fluid model can be compared with those experimental results, as shown in table 2. Since different modes of nanoparticles have been experimentally excited under proper external stimuli, the comparison in table 2 includes the fundamental or higher-order breathing ($n=0$), dipolar ($n=1$), and quadrupolar ($n=2$) modes. It is found that the theoretical predictions based on the SSM agree quite well with the experimental results.

In a word, we can use the present SSM and Muller's method to obtain accurate numerical results with a high precision. In the following calculations, the number of the equally discretized thin layers is taken to be 50 for which the numerical results can be assumed to be highly accurate. If not otherwise stated, a gold nanosphere with radius $b = r_N$ immersed in different glycerol-water mixtures is considered and the associated material properties are all provided in SM-III for completeness. In addition, in what follows, we will only pay attention to the smallest nonzero root (i.e., the first-order radial mode) for each angular mode $n$ and the higher order ones can be readily obtained if needed.

### 5.2. Effects of fluid viscosity and compressibility

In this subsection, the vibration characteristics in vacuum, non-viscous incompressible, non-viscous compressible, viscous incompressible, and viscous compressible fluids will be shown separately for comparison and illustration.

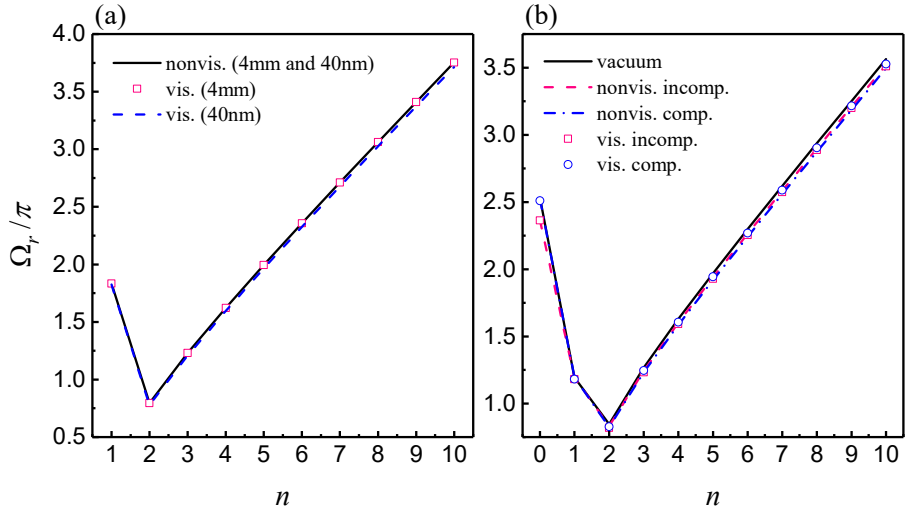

FIGURE 2. Normalized vibration frequencies versus the angular mode number $n$ of an isotropic gold sphere in a glycerol-water mixture ($\chi = 0.56$) using the Newtonian model: (a) torsional modes for two different radii 4mm and 40nm; (b) spheroidal modes for 40nm.

The effects of fluid viscosity and compressibility on the normalized vibration frequencies $\Omega_r / \pi$ and damping components $\Omega_i / \pi$ are depicted in figures 2 and 3, respectively, for the two classes of vibration of an isotropic solid sphere submerged in a glycerol-water mixture modelled by the Newtonian fluid. Note that the vibration attenuation originates from two aspects: energy dissipation due to the viscosity of the fluid and outgoing energy propagation because of fluid compressibility (Ding & Chen 1998; Galstyan, Pak & Stone 2015). Since the first class of vibration is an equi-volumetric motion, the fluid compressibility does not affect the free vibration characteristics of the sphere. Therefore, figures 2(a) and 3(a) only show the effect of fluid viscosity. One can observe from figure 2 that for the first class of vibration, the vibration frequency for the purely torsional or rotary mode ($n=1$) is higher than that of the mode ($n=2$), and the frequency of the torsional mode increases with the increasing angular mode number when $n \geq 2$; the vibration frequencies of the second class corresponding to the non-breathing modes ($n = 1 \sim 6$) are lower than that of the breathing mode $n = 0$. As mentioned before, different vibration modes of nanoparticles have been excited and observed in experiments (Liu *et al.* 2009; Crut *et al.* 2015), and hence it is extremely important to theoretically study the vibration characteristics of general vibration modes other than the simple breathing mode.

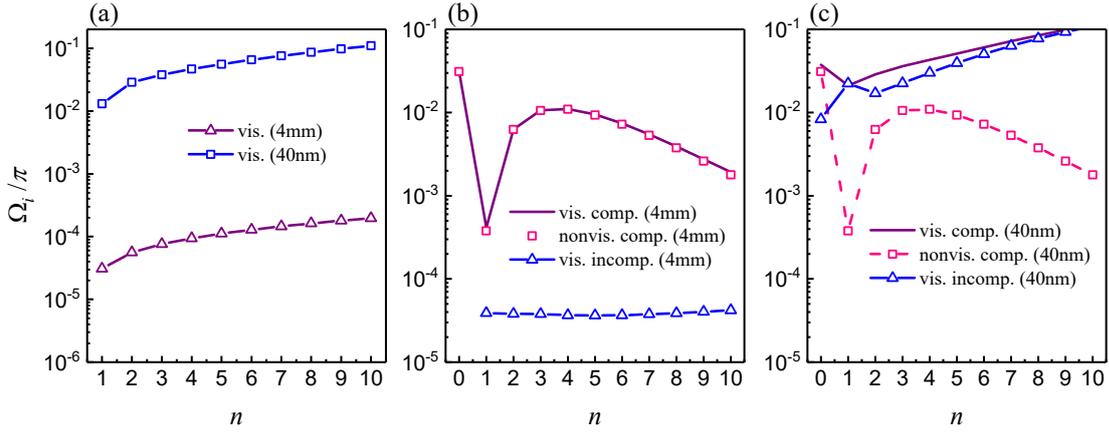

FIGURE 3. Normalized damping components versus the angular mode number *n* of an isotropic gold sphere in glycerol-water mixture ($\chi = 0.56$) using the Newtonian model: (a) torsional modes for 4mm and 40nm; (b, c) spheroidal modes for 4mm and 40nm.

The results in figures 2 and 3 show that fluid compressibility and viscosity have little effect on the resonant frequency of the two classes of vibration, but play a more obvious role in the damping component. This indicates that the free vibration frequency is hardly affected by the properties of the surrounding fluid while the attenuation behavior is strongly dependent on the fluid compressibility and viscosity. Specifically, for the first class of vibration, the damping component increases as the mode number increases and the radius of the sphere decreases, which means that the effect of viscosity on the attenuation of higher-order angular modes and smaller nano-sized spheres is more significant than that of lower-order angular modes and larger spheres (say at a macroscale). For the second class of vibration, it can be seen from figure 3(b) that, for a macroscopic dimension ($b = 4\,\text{mm}$), the effect of viscosity is small and the attenuation is mainly fixed by the fluid compressibility. The attenuation of the higher-order modes is smaller than that of the breathing mode. For the nanosphere ($b = 40\,\text{nm}$) shown in figure 3(c), the result corresponding to the non-viscous compressible case is hardly changed compared with that of a macroscopic size in figure 3(b). Nonetheless, the effect of viscosity becomes more pronounced, eventually leading to an increase in the attenuation of the second class of vibration over the entire mode range of interest.

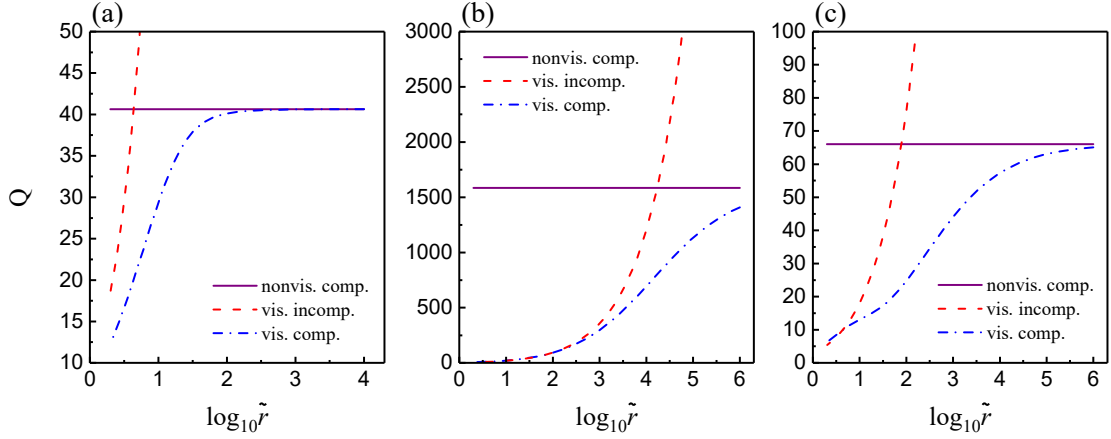

FIGURE 4. Quality factors of the second class of vibration versus the logarithmic quasi Reynolds number $\log_{10}\tilde{r}$ of an isotropic gold sphere in glycerol-water mixture ($\chi=0.56$) using the Newtonian model for the breathing $n=0$ (a), dipolar $n=1$ (b), quadrupolar $n=2$ (c) modes. The quasi Reynolds number is defined as $\tilde{r}=\rho_f c_f b/\eta$.

In order to clearly show the dependence of vibration characteristics on the size of the sphere and the fluid viscosity, we display the variations of the quality factor of the submerged isotropic gold sphere with the logarithmic quasi Reynolds number $\log_{10}\tilde{r}$ in figure 4 for the breathing ($n=0$), dipolar ($n=1$), and quadrupolar modes ($n=2$) of the second class. It is interesting to note that for a non-viscous compressible fluid, the damping of which is solely due to the propagation of acoustic waves away from the vibrating sphere, the quality factor does not change with the quasi Reynolds number $\tilde{r}$, which means that there is no size effect. Nevertheless, when viscosity effect is considered, the vibration characteristics vary with $\tilde{r}$ exhibiting the size effect. In fact, in addition to the radius $b$ of the solid sphere, there exists another intrinsic size $\eta/\rho_f c_f$ of the fluid due to the presence of fluid viscosity. This is the right reason of the appearance of the size-dependent characteristics. Specifically, a general variation trend that the quality factors of all vibration modes of the sphere in the viscous fluid go down with the decrease of $\tilde{r}$ is observed in figure 4. Since the density and sound speed do not alter significantly for different fluids, the reduction of the quasi Reynolds number $\tilde{r}=\rho_f c_f b/\eta$ can be classified into two cases, namely, the decrease of the radius $b$ of the sphere and the increase of the fluid viscosity $\eta$. As a result, when $\tilde{r}$ is infinitely large, the attenuation induced by fluid viscosity becomes small, the attenuation behavior mainly relies on fluid compressibility, and the vibration quality factor for the viscous compressible fluid approaches the non-viscous compressible result. However, when $\tilde{r}$ reduces gradually, the viscosity-induced attenuation has a dramatic increase. In this case, the damping component is mainly determined by the viscosity, and thus the quality factor asymptotically tends towards the viscous incompressible result (see figure 4), which has a limiting value of 0.5. In short, when the size of the vibrating sphere decreases into the nanometer scale or the shear viscosity of the fluid increases, fluid viscosity has a significant influence on the vibration characteristics of the FSI system and must be taken into account (Galstyan, Pak & Stone 2015).

For the first class of vibration, numerical calculations have also been conducted for the submerged gold sphere for different torsional modes and glycerol-water mixtures. The results are qualitatively similar to those for the second class of vibration (see figure S2 in SM-V).

### 5.3. *Effect of fluid viscoelasticity*

Now we turn to elucidating the effect of fluid viscoelasticity on the vibration characteristics of the submerged nanosphere. When viscoelasticity presents, the Deborah number $De = \lambda \omega$ is usually adopted (Chakraborty & Sader 2015; Chakraborty *et al.* 2018), where $\lambda$ and $\omega$ represent the characteristic molecular relaxation time and vibration circular frequency of the structure. Nonetheless, since $\omega$ (especially the damping component) strongly depends on the viscoelastic properties of the surrounding fluid, we will scale the shear and compressional relaxation times $\lambda_{sh}$ and $\lambda_{comp}$ by the circular frequency $\omega_{vac}$ of the structure in vacuum and thus define two *quasi Deborah numbers* $De_{sh} = \lambda_{sh} \omega_{vac}$ and $De_{comp} = \lambda_{comp} \omega_{vac}$. Note that $\omega_{vac}$ is only determined by the structure itself and not affected by the fluid.

Based on the linear Maxwell ($De_{comp} = 0$) and compressional non-Newtonian ($De_{comp} \neq 0$) models, the variations with the logarithmic quasi Reynolds number $\log_{10} \tilde{r}$ of the normalized vibration frequency $\omega_r / \omega_{vac}$ and the logarithmic quality factor $\log_{10} Q$ of a submerged sphere are depicted in figure 5 for the breathing mode and different quasi Deborah numbers. When the compressional non-Newtonian model is employed, the condition $De = De_{sh} = De_{comp}$ is taken as a good approximation for some real fluids (Chakraborty & Sader 2015; Chakraborty *et al.* 2018). It can be seen from figure 5 that both the vibration frequency and quality factor predicted by these two models for all quasi Deborah numbers asymptotically tend towards the non-viscous compressible results for large $\tilde{r}$ (small viscosity effect) and approach the incompressible viscoelastic results for small $\tilde{r}$ due to the viscoelasticity-dominated influence (Chakraborty & Sader 2015). The quality factor corresponding to the Newtonian model ($De = 0$) shown in figure 5(b) declines monotonically with the decreasing $\tilde{r}$, which is similar to the result in figure 4(a). However, the vibration frequency for $De = 0$ in figure 5(a) exhibits a non-monotonic behavior as a function of $\tilde{r}$, which may be attributed to the complex mutual interaction of fluid viscosity, compressibility, and structural size (Chakraborty & Sader 2015).

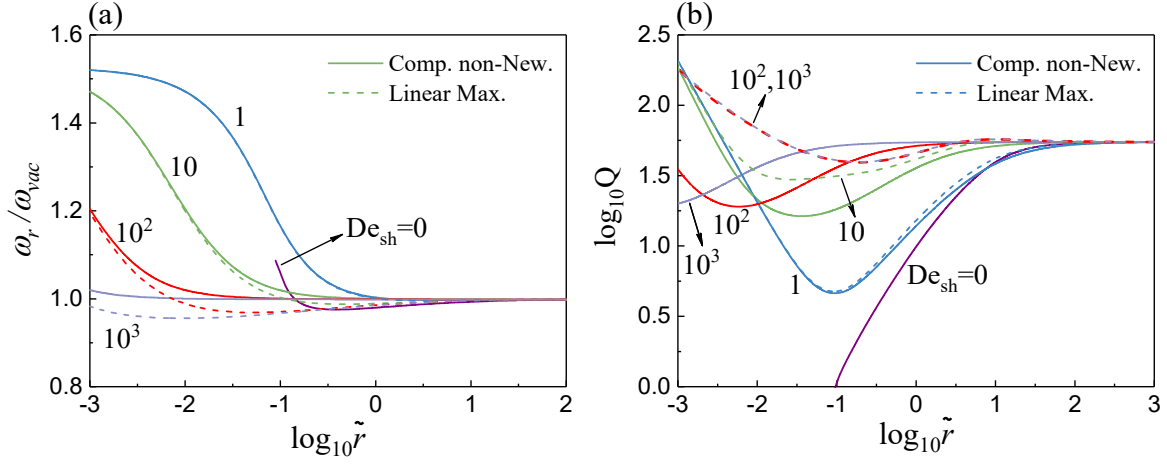

FIGURE 5. Normalized vibration frequency (a) and logarithmic quality factor (b) of the breathing mode ($n=0$) versus the logarithmic quasi Reynolds number $\log_{10}\tilde{r}$ for an isotropic sphere immersed in a compressible viscoelastic fluid with different quasi Deborah numbers $\mathrm{De}_{sh}=0$ (compressible Newtonian model), 1, 10, $10^2$, and $10^3$. Results are provided based on the linear Maxwell model (dashed lines) and compressional non-Newtonian model (solid lines) for the dimensionless quantities $\tilde{\rho}=0.05$, $\tilde{v}=0.8$, $\gamma=1$, and Poisson's ratio $\nu=0.42$. The quasi Deborah numbers are chosen as $\mathrm{De}_{comp}=\mathrm{De}_{sh}$ for compressional non-Newtonian model.

Furthermore, results in figure 5 display that the predictions based on the linear Maxwell and compressional non-Newtonian models for finite quasi Deborah numbers (i.e., $\mathrm{De}\neq 0$) demonstrate a dramatical difference from the Newtonian results. Specifically, the vibration frequency increases remarkably with a reduction in $\tilde{r}$, especially in the low $\tilde{r}$-region. Additionally, instead of a reduction in quality factor with decreasing $\tilde{r}$, the quality factor has a monotonically reverse increase after a critical value $\tilde{r}_c$ is reached. These seemingly unintuitive behaviors originate from the significant effect of fluid elasticity: the stored energy in the FSI system overly compensates the energy dissipation induced by fluid viscosity and compressibility, which results in the remarkable rise in the vibration frequency and quality factor in figure 5 and will be discussed further below for a real FSI system (i.e., a gold nanosphere vibrating in a glycerol-water mixture).

Moreover, it is obvious from figure 5 that for a fixed $\tilde{r}$ of a moderate or small magnitude, the vibration frequency provided by the compressional non-Newtonian model is larger than that predicted by the linear Maxwell model, while the quality factor shows an opposite trend. The reason for this phenomenon can be explained as follows. For the breathing mode with the nature of compressional motion, the compressional elasticity stiffens the FSI system, leading to the increase of vibration frequency. Nonetheless, the compressional relaxation effect enhances the compressibility effect and thus increases the energy radiation (the calculated damping component of the circular frequency by the compressional non-Newtonian model is larger than that based on the linear Maxwell model), producing a decrease in the quality factor of the breathing mode, especially for high quasi Deborah numbers. This phenomenon will be also investigated further below for a specific FSI system by means of spatial distribution of the stored energy density.

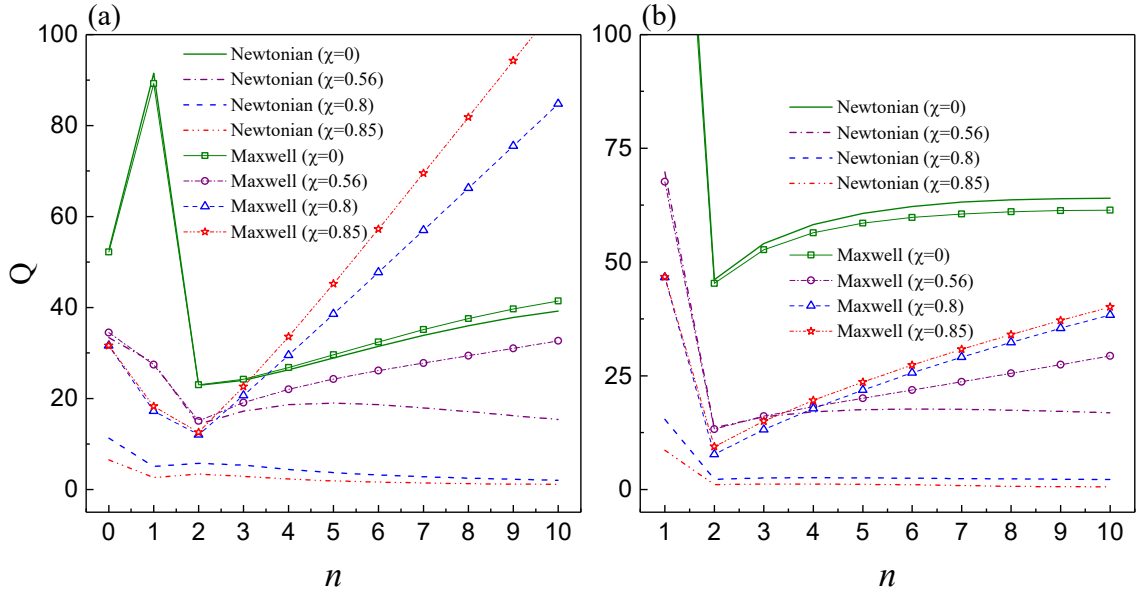

FIGURE 6. Variations of the quality factor with the angular mode number $n$ of a 40nm isotropic gold nanosphere for different glycerol-water mixtures using both Newtonian and linear Maxwell models: (a) spheroidal modes; (b) torsional modes.

To highlight the difference between the Newtonian and non-Newtonian fluid models, figure 6 shows the variation of the quality factor of two classes of vibration with the angular mode number for a 40nm isotropic gold nanosphere in different glycerol-water mixtures. It can be seen that for the two classes of vibration and low glycerol mass fraction $\chi$ (i.e., the effect of viscoelasticity is small), the predictions based on the two models are not much different when the mode is lower than a certain critical mode. For instance, the critical modes for the second class of vibration corresponding to $\chi = 0$ and $\chi = 0.56$ are $n = 6$ and $n = 3$, as shown in figure 6(a). Beyond the critical mode, the difference between these two models increases gradually with the increasing angular mode number. It is worth pointing out that the critical mode number varies and depends on the size and glycerol mass fraction, which will be discussed in detail below. Furthermore, for high glycerol mass fractions such as $\chi = 0.8$ and $\chi = 0.85$, the obtained results based on the two models are extremely different over the entire mode range, especially for the higher-order mode numbers. Specifically, as the mode number increases, the quality factor predicted by the Newtonian model monotonically decreases to a limiting value of 0.5, while that based on the linear Maxwell model first decreases and then increases reversely.

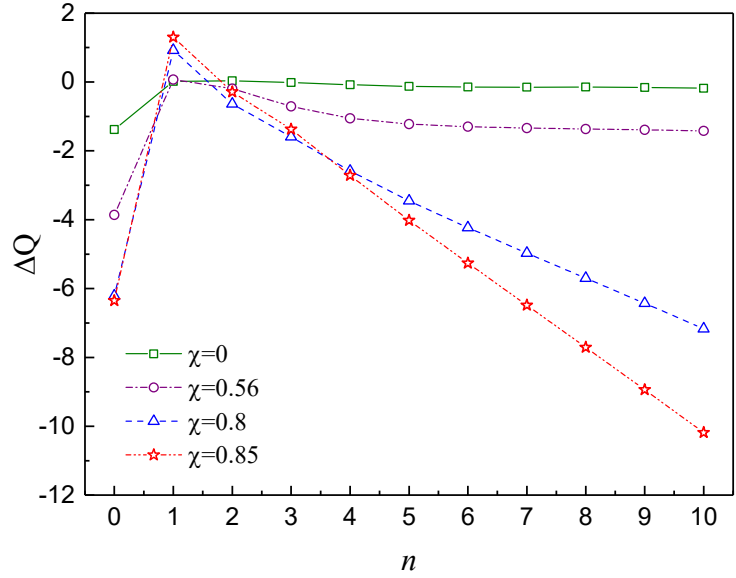

FIGURE 7. Variations of the quality factor difference $\Delta Q = Q_1 - Q_2$ with the angular mode number $n$ of a 40nm isotropic gold nanosphere for different glycerol-water mixtures using two different non-Newtonian models, where $Q_1$ is the quality factor predicted based on the compressional non-Newtonian model, while $Q_2$ is that by the linear Maxwell model.

The variation trend predicted by the compressional non-Newtonian model accounting for both the shear and compressional relaxation effects is qualitatively consistent with that of the linear Maxwell model, and hence it is not shown here for brevity. Instead, we will only demonstrate the difference of the predictions based on the two fluid models. Note that for the first class of vibration, the compressional relaxation effect does not alter the free vibration characteristics of the submerged sphere due to its equi-volumetric nature. However, for the second class of vibration, the compressional relaxation process indeed affects the vibration frequency and quality factor. The difference of the quality factor of the second class of vibration predicted by these two fluid models is plotted in figure 7 for different glycerol mass fractions. It is found that for low mass fractions, the compressional relaxation still has a noticeable influence on the breathing mode $n = 0$ because of its compressional motion behavior, but its effect on the higher-order modes is less significant compared with the breathing mode and is negligible. For high mass fractions, the vibration characteristics strongly depend upon the compressional relaxation process over the whole mode range.

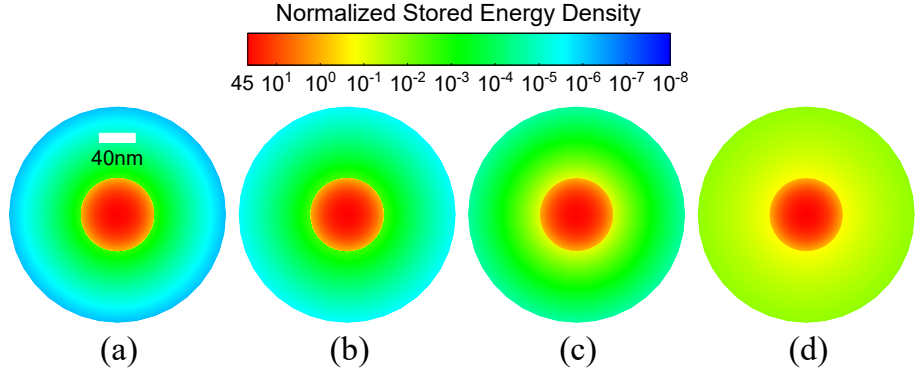

FIGURE 8. Spatial distributions of the normalized exceptional maximum stored energy density $E^{\text{exc}}$ in a vibrating isotropic gold nanosphere ($b = 40\,\text{nm}$) and in the surrounding glycerol-water mixture for the breathing mode $n = 0$ using the linear Maxwell model (a, c) and the compressional non-Newtonian model (b, d): (a, b) $\chi = 0$ (pure water); (c, d) $\chi = 0.8$. The color scale represents the exceptional maximum stored energy density normalized by the energy density at the outer surface of the nanosphere. The radial thickness of the fluid computational domain is limited within $2b$.

Moreover, for different glycerol-water mixtures, figure 8 depicts the spatial distributions of the *normalized exceptional maximum stored energy density* $E^{\text{exc}}$ (see SM-IV for more details) in the vibrating gold nanosphere as well as in the surrounding fluid for the breathing mode $n = 0$ predicted by the linear Maxwell model and the compressional non-Newtonian model. Since the breathing mode has only the radial displacement component (which is independent of $\theta$ and $\varphi$), the spatial distribution of the energy density only varies along the $r$-direction and does not change along the $\theta$-direction, as shown in figure 8. One can observe that the compressional relaxation effect on the energy density distribution of the breathing mode is more significant for a high glycerol mass fraction ($\chi = 0.8$) than a low mass fraction ($\chi = 0$). In particular, since the breathing mode is a spherically symmetric compressional motion, the compressional relaxation enhances the outward energy propagation due to the vibrating nanosphere and increases the energy attenuation, leading to a decrease in the quality factor of the breathing mode (see figure 7). Overall, for complex fluids with considerable viscoelastic effect and for the compression-dominated flow induced by compressional motion, it is necessary to use the compressional non-Newtonian model instead of the Newtonian and linear Maxwell models to accurately predict the vibration characteristics.

Now the compressional non-Newtonian model is utilized to explore under what conditions the viscoelastic effect should be taken into consideration. For different vibration modes and different radii of the nanosphere, figure 9 displays the quality factors $Q$ predicted by both the Newtonian and compressional non-Newtonian models as a function of the glycerol mass fraction $\chi$ for the second class of vibration. It is found that the quality factor based on the Newtonian model monotonically decreases to the limiting value of 0.5 with the increase of $\chi$. However, as the mass fraction increases to a certain critical value $\chi_c$, the quality factor given by the compressional non-Newtonian model no longer decreases, but remains almost unchanged (e.g. the breathing mode $n=0$), and even goes up reversely for some modes (e.g. the quadrupolar mode $n=2$ and the mode $n=6$). In fact, as Pelton *et al.* (2013) pointed out, for low mass fractions with weak fluid elastic effect, the dissipation due to viscosity and compressibility results in the decrease of the quality factor and the results for both models agree with each other quite well. For mass fractions higher than the critical value $\chi_c$, the stored energy in the fluid can compensate or even exceed the dissipated energy induced by viscosity and compressibility. Therefore, the quality factor no longer declines and can even increase with the increasing glycerol mass fraction, which was also observed by Galstyan, Pak & Stone (2015) for the breathing mode of a gold nanosphere vibrating in a linear Maxwell fluid. Essentially the same analysis can be conducted for torsional modes of the first class of vibration (see figure S3 in SM-V).

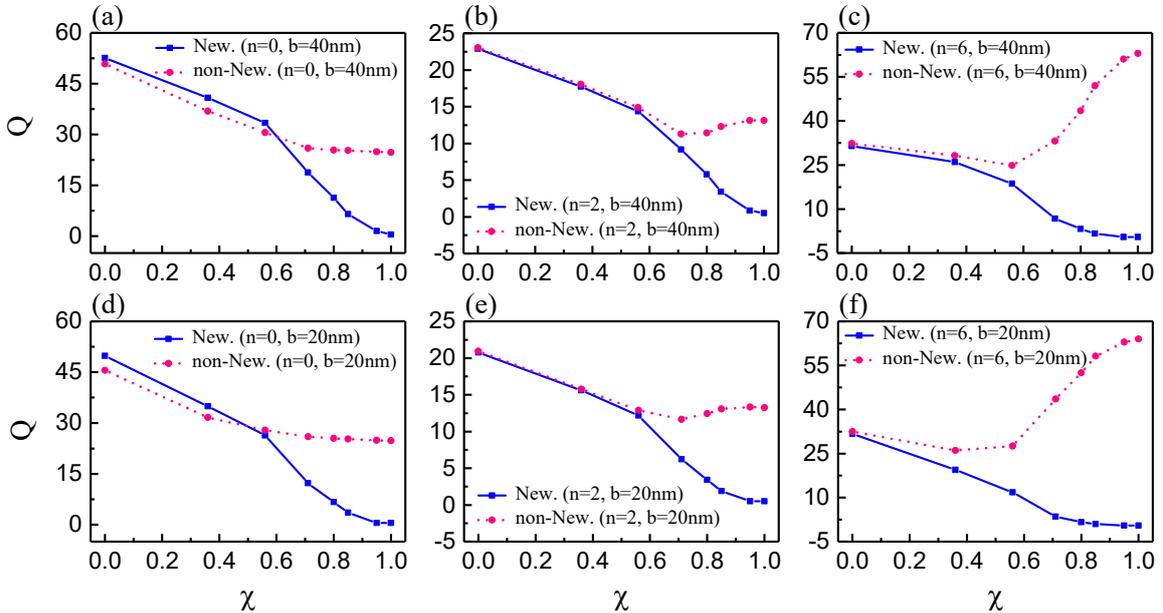

FIGURE 9. Quality factor $Q$ of the second class of vibration as a function of the glycerol mass fraction $\chi$ for the isotropic gold nanosphere with different radii $b=40\,\mathrm{nm}$ (a-c) and $b=20\,\mathrm{nm}$ (d-f) and different spheroidal modes $n=0$ (a, d), $n=2$ (b, e), and $n=6$ (c, f) using both the Newtonian and compressional non-Newtonian models.

In addition, one can observe from figure 9 that the critical glycerol mass fraction $\chi_c$ depends on the mode number and the radius of the nanosphere. In fact, Eq. (4) demonstrates that the elastic effect of the fluid will become more obvious if $|\omega\lambda_{sh}|\geq 1$ or $|\omega\lambda_{comp}|\geq 1$, i.e., when the relaxation time scale of the fluid $\lambda_{sh}$ or $\lambda_{comp}$ is comparable to the vibration time scale $1/\omega_r$ (Galstyan, Pak & Stone 2015). Based on the given vibration frequency $\omega_r$, one can therefore estimate the critical glycerol mass fraction $\chi_c$ beyond which the prediction by the compressional non-Newtonian model will deviate considerably from that by the Newtonian model. For example, for the second class of vibration of a 40nm gold nanosphere, the vibration frequencies $f=\omega_r/2\pi$ corresponding to the spheroidal modes $n=0$, 2 and 6 in vacuum are calculated as $f\approx 38\,\text{GHz}$, $13\,\text{GHz}$ and $35\,\text{GHz}$, respectively. Note that it can be numerically confirmed that the vibration frequency is not significantly affected by the surrounding fluid considered here. The critical relaxation times determined from the condition $|\omega\lambda_c|=1$ are approximately $\lambda_c\approx 4.2\,\text{ps}$, $12.2\,\text{ps}$ and $4.5\,\text{ps}$, which correspond to the critical mass fractions $\chi_c$ in the range of $0.5\sim 0.7$ (referring to table S1 in SM for the shear and compressional relaxation times of different glycerol-water mixtures). These are consistent with the results in figure 9. We can also execute similar analysis for the torsional modes of the first class of vibration (see figure S3 in SM-V). Interestingly, for the spheroidal mode $n=6$ with $b=20\,\text{nm}$ (see figure 9(c)) and for the torsional mode $n=1$ with $b=10\,\text{nm}$ (see figure S3(c) in SM-V), the deviation of the predictions based on the two models shows up even when the glycerol mass fraction is close to zero (i.e. pure water), indicating that for higher-order angular modes or smaller spherical radii with high vibration frequencies, the viscoelastic effect on fluid-solid coupled vibrations becomes significant even in a low-viscosity fluid. As a consequence, the FSI vibration characteristics at nanoscale can be exploited to measure the viscoelastic properties of low-viscosity fluids (such as pure water) with faster relaxation times.

In order to evidently demonstrate the stored energy evolution with the glycerol mass fraction, the spatial distributions of $E^{exc}$ in a 40nm isotropic nanosphere and different surrounding mixtures are illustrated in figure 10 for the breathing ($n=0$) and quadrupolar ($n=2$) modes of the second class of vibration using the compressional non-Newtonian model. Note that for axisymmetric modes (i.e. $m=0$), it is only necessary to give the energy distribution in the $r-\theta$ plane. For the breathing mode, the energy distribution does not change along the $\theta$-direction, while for the quadrupolar mode, the displacement and stress components depend on $\theta$ resulting in the variation of the energy distribution with $\theta$. However, the energy distribution of the quadrupolar mode is still symmetric, as shown in figures 10(d-f). It can be seen that as expected, more energy is stored in the surrounding fluid due to the increasing elastic effect of the fluid as the glycerol mass fraction increases, which can compensate or even exceed the dissipated energy. The overall stiffness of the FSI system also increases with the increasing glycerol mass fraction, which ultimately leads to the increase in the vibration frequency (not shown here for simplicity).

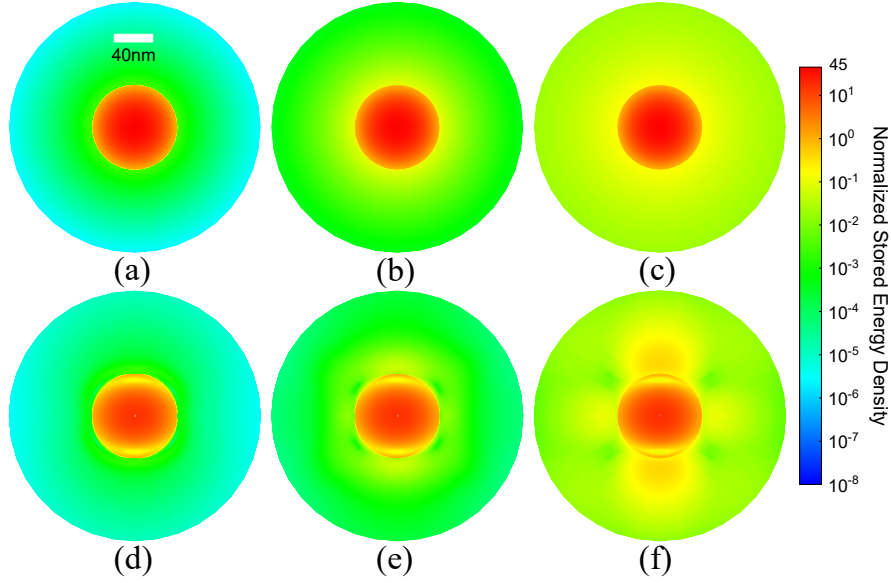

FIGURE 10. Spatial distributions of the normalized exceptional maximum stored energy density $E^{exc}$ of the second class of vibration in a 40nm isotropic gold nanosphere and in the surrounding glycerol-water mixture for the breathing mode $n = 0$ (a-c) and quadrupolar mode $n = 2$ (d-f) using the compressional non-Newtonian model: (a, d) $\chi = 0$ (pure water); (b, e) $\chi = 0.56$; (c, f) $\chi = 0.85$. The color scale represents the exceptional maximum stored energy density normalized by the energy density at the north pole of the nanosphere for each mode and mass fraction. The radial thickness of the fluid computational domain is limited within $2b$.

For different glycerol mass fractions and different vibration modes, the variations of the quality factor of the second classes of vibration with the logarithmic spherical radius $\log_{10}[b(\text{nm})]$ are depicted in figure 11 using the Newtonian and compressional non-Newtonian fluid models. As mentioned above, the quality factor predicted by the Newtonian model monotonically decreases with the decreasing radius in all cases because of the increasing viscous dissipation effect. In contrast, as the radius decreases from macroscale to nanoscale, the quality factor based on the compressional non-Newtonian model first decreases to a minimum, and then remains almost unchanged or even has an opposite rise after the radius decreases to a certain critical value $b_c$. When the radius is larger than $b_c$, the predictions of the two models are almost the same since the vibration frequency is not high enough for the elastic effect of the fluid to play a role. For a macroscopic radius, the vibration characteristics approach the non-viscous compressible results. On the other hand, if the radius is smaller than $b_c$, the extremely high vibration frequency enhances the elastic effect of the glycerol-water mixture, and hence the quality factor no longer decreases and can even increase with the decreasing radius (see figure 11).

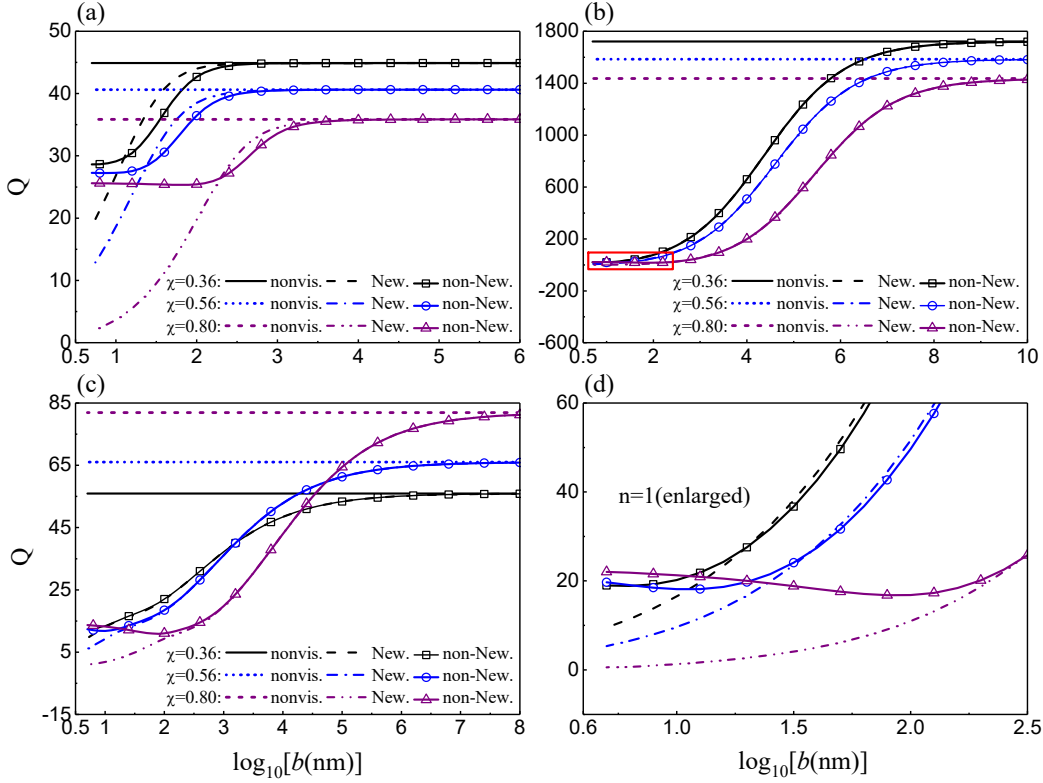

FIGURE 11. Quality factor $Q$ of the second class of vibration as a function of the logarithmic radius $\log_{10}[b(\text{nm})]$ for the isotropic gold nanosphere in different glycerol-water mixtures using both the Newtonian and compressional non-Newtonian models for different spheroidal modes: (a) breathing mode $n=0$; (b) dipolar mode $n=1$; (c) quadrupolar mode $n=2$; (d) $n=1$ (enlarged).

The critical radius $b_c$ depends on the glycerol mass fraction as well as mode number and it can be estimated using the relation $|\omega_c \lambda| = 1$ as described earlier. For instance, for the breathing mode in figure 11(a), the critical vibration frequencies $f_c = \omega_{rc}/2\pi$ corresponding to glycerol mass fractions $\chi = 0.8$, 0.56, and 0.36 can be calculated as $f_c \approx 5.8\text{GHz}$, 43GHz and 85GHz, respectively, according to the relaxation times listed in table S1 in SM. Utilizing the numerical results of the vibration frequency of gold nanospheres in vacuum for the breathing mode, the corresponding critical radii of the sphere are $b_c \approx 260\text{nm}$, 29nm and 17nm, which occur when $\log_{10}[b_c(\text{nm})] \approx 2.41$, 1.46 and 1.23, respectively, as displayed in figure 11(a). The discussions can be similarly applied to the dipolar and quadrupolar modes as well as the torsional modes of the first class of vibration (see figure S4 in SM-V). To clearly demonstrate the size effect of the spherical particle on energy distributions, figure S5 in SM-V further shows the spatial distributions of $E^{\text{exc}}$ in the vibrating gold nanosphere with different radii and in the surrounding fluid for the breathing mode. Qualitatively similar physical phenomena to those in figure 11 can be observed. Thus, we can exploit this behavior to properly select the size and vibration mode of nanostructures to measure the viscoelastic properties of a fluid with a given mass fraction.

### 5.4. *Effect of solid surface layer*

In general, single crystal (or monocrystalline) nanoparticles are not isotropic and exhibit various degree of anisotropy in material properties. To save space of the paper, we demonstrate in SM-V-C the capability of the previously derived frequency equations (39)-(42) in studying the anisotropy effect of the nanosphere on the dynamic properties of the FSI system. The interested reader is therefore referred to SM-V-C for more details.

At nanoscale, it is well known that the atomic structure in or near the surface of a medium usually experiences a different local environment from that in the bulk counterpart, such that the material properties of the surface are distinct from those of the bulk (Cammarata 1997). In addition, a nanoscale surface adhesion layer with a certain thickness can be artificially utilized to actively control the vibration behavior of the nanostructures (Chang et al. 2015). Therefore, we consider here a concentric core-shell structure with different material properties in the core and shell. This core-shell model has been used to investigate the surface effect on the dynamic properties of nanostructures in vacuum (Chen *et al.* 2014; Wu et al. 2015; Wu, Chen & Zhang 2018). Now our interest is to employ the developed multilayered spherical model to examine how the surface effect of the nanosphere influences the vibration characteristics of the FSI system. The core-shell nanosphere is assumed to be cubic and has a spherical solid core of radius $b$ as well as a thin surface layer of thickness $h$ which can be approximately taken as 1 nm in our calculations. Two dimensionless elastic and density parameters ($r_c = c_{ij}^{(2)}/c_{ij}^{(1)}$ and $r_\rho = \rho_s^{(2)}/\rho_s^{(1)}$) are adopted to study the surface effect for different surface material properties $c_{ij}^{(2)}$ and $\rho_s^{(2)}$, wherein $c_{ij}^{(1)}$ and $\rho_s^{(1)}$ denote the material properties of the spherical core.

For the breathing mode of a 20nm cubic gold nanosphere ($r_\rho = 1$), the variations of the normalized vibration frequency, damping component, and quality factor with the glycerol mass fraction are illustrated in figure 12 for different surface elastic parameters using the compressional non-Newtonian fluid model. It can be found that the variation trend with the surface effect ($r_c \neq 1$) is qualitatively similar to that without surface effect ($r_c = 1$). With the increase of the surface elastic parameter, the vibration frequency of the breathing mode increases over the entire glycerol mass fraction range, but the damping component has an opposite variation trend, thus resulting in the rise of the quality factor. In fact, an increase in surface elasticity enhances the overall stiffness of the FSI system, makes more energy be stored in the nanosphere, and hence raises the vibration frequency and quality factor. We also conduct the numerical analysis on the effect of surface density parameter on the FSI vibration characteristics (see SM-V-D for more details). The increasing surface mass density greatly weakens the effect of the surrounding fluid on the damping component and hence enhances the vibration quality factor, which is physically quite understandable.

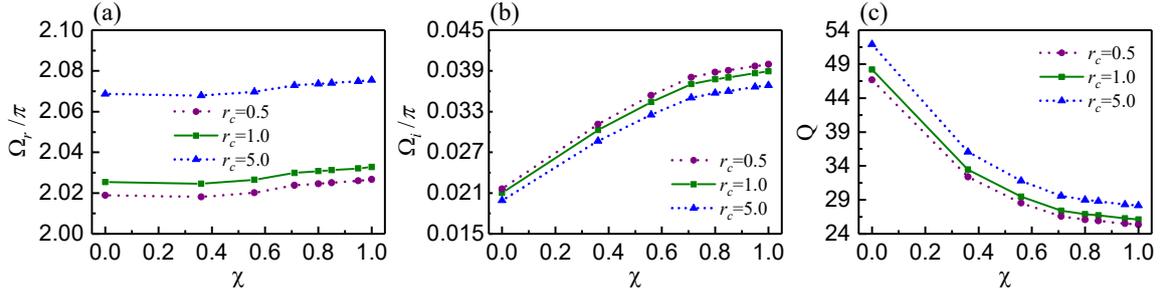

FIGURE 12. Normalized vibration frequency (a), damping component (b), and quality factor (c) of the breathing mode $n = 0$ as functions of the glycerol mass fraction for a 20nm cubic gold nanosphere with surface effect ($r_\rho = 1$ and different surface elastic parameters $r_c$) using the compressional non-Newtonian model.

Now let us turn our attention to the surface effect on the variation trend of the vibration characteristics with the size of the nanosphere. For the breathing mode $n = 0$ of a cubic gold nanosphere in a glycerol-water mixture ($\chi = 0.56$), the variations of the normalized vibration frequency, damping component, and quality factor with the logarithmic radius $\log_{10}[b(\text{nm})]$ are illustrated in figure 13 using the compressional non-Newtonian model for different surface elastic and mass density parameters. It can be seen that when there is no surface effect, the vibration characteristics tends towards the non-viscous compressible results for an infinitely large radius; with the radius decreasing, the vibration frequency hardly changes as described earlier, while the damping component gradually increases to a constant value because of the increasing viscoelastic effect, thus causing the quality factor to first decrease and then approach a stable value. When surface effect is involved, the vibration characteristics for a macroscopic size tend towards the results without surface effect, which is expected since the surface effect becomes negligible for a macroscopic sphere. However, when the radius of the sphere reduces to a critical value, the vibration characteristics begin to deviate from the results without surface effect. For instance, the critical radii for quality factors in figures 13(c) and 13(f) are approximately equal to 32nm and 56nm, respectively.

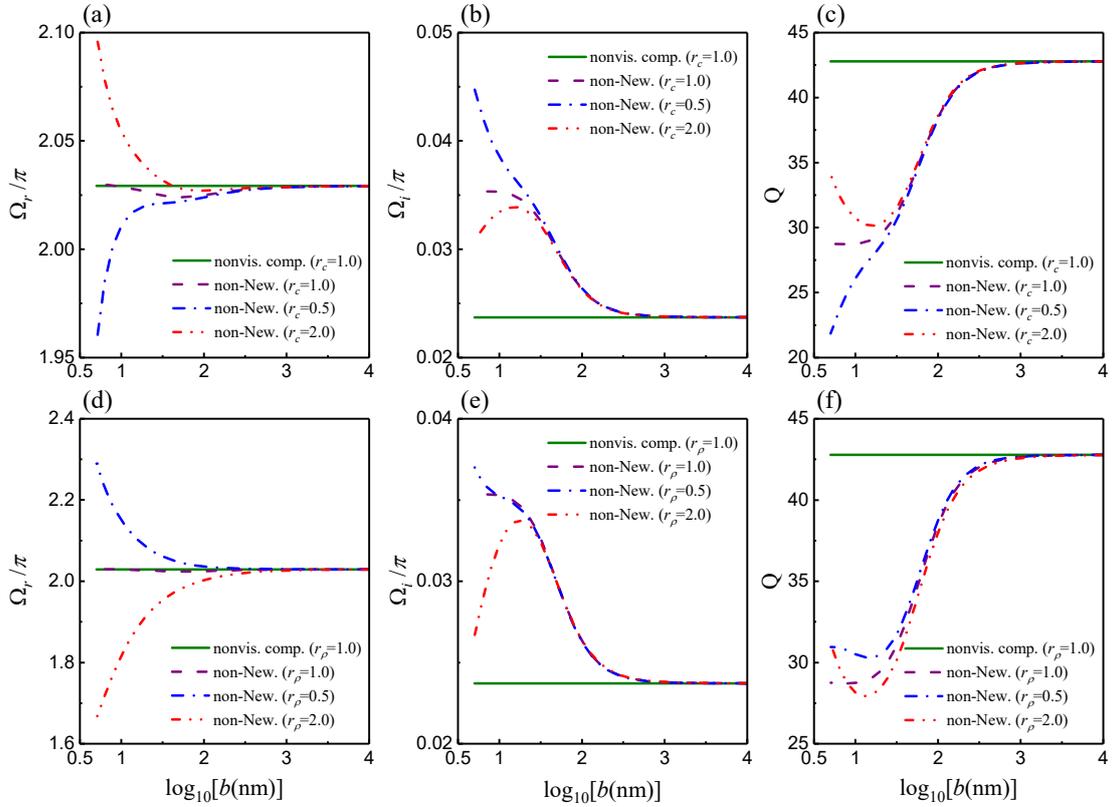

FIGURE 13. Normalized vibration frequencies (a, d), damping components (b, e), and quality factors (c, f) of the breathing mode $n=0$ of a cubic gold nanosphere with surface effect in a glycerol-water mixture ($\chi=0.56$) as functions of the logarithmic radius using the compressional non-Newtonian model: (a-c) $r_\rho=1$ and different surface elastic parameters; (d-f) $r_c=1$ and different surface density parameters.

Furthermore, the results in figures 13(a-c) show that if the surface layer becomes softer ($r_c=0.5$), the vibration frequency decreases but the damping component further increases due to the increasing viscous dissipation, leading to a continuous reduction of the quality factor with the decreasing radius. Nonetheless, for a stiffer surface layer ($r_c=2$), as the radius decreases, the vibration frequency continuously increases while the damping component first increases and has an opposite decrease after reaching a maximum, which makes the quality factor go up remarkably after the radius reaches the critical value. In fact, the stiffer surface layer makes more elastic energy stored in the nanosphere which exceeds the dissipated energy induced by fluid viscosity and compressibility, giving rise to the obvious reverse increase of the quality factor. Again, the heavier surface layer reduces the effect of fluid inertia on the vibration characteristics of the nanosphere, resulting in a significant raise of quality factor (see figure 13(f)).

Therefore, when the surface layer is properly chosen in terms of stiffness and density, the quality factor is prone to achieving an opposite increase with the decreasing nanosphere size for the breathing mode. To conclude, we can actively modulate the vibration characteristics of the nanosphere submerged in a viscoelastic fluid by surface engineering. Additionally, we can also exploit the surface-dependent nanoscale FSI vibrations to design new molecular and atomic sensors with high sensitivity (Pelton *et al.* 2013).

### 5.5. *Effect of solid intrinsic damping*

It is worth pointing out that the aforementioned theoretical and numerical investigations only take account of the radiation and viscous damping resulting from fluids, and assume that the nanospheres are purely linear elastic without damping. In practice, different nanoparticles used in experiments may be equipped with some intrinsic damping mechanisms such as internal defects and dissipation in the capping layer surrounding the nanoparticle. This could be the reason leading to the discrepancy between the experimentally measured and theoretically predicted vibration quality factors (Pelton *et al.* 2011; Ruijgrok *et al.* 2012; Chakraborty *et al.* 2013).

When solid intrinsic damping is taken into account, the elastic tensor is no longer real but also has a complex part (i.e., the real elastic stiffness plus an imaginary damping stiffness). In this work, we will employ the *hysteretic* viscoelastic model (Quintanilla *et al.* 2015). Specifically, the complex Young's modulus $E_Y$ can be written as $E_Y = E_r - \mathrm{i}E_i$, where $E_r$ specifies the storage elastic modulus while $E_i$ indicates the loss modulus due to the damping. Here, $E_i$ is assumed to be frequency independent. To consider frequency-dependent loss modulus, other linear viscoelastic models such as the Kelvin-Voigt model may be adopted (Quintanilla *et al.* 2015). The quality factor only including the contribution from fluid dissipation and taking no account of solid intrinsic damping is referred to as fluid quality factor $Q_{\text{fluid}}$, while $Q_{\text{intrinsic}}$ indicates the quality factor of nanoparticles with intrinsic damping in vacuum. The total quality factor $Q_{\text{total}}$ accounts for all damping mechanisms in both fluids and solids.

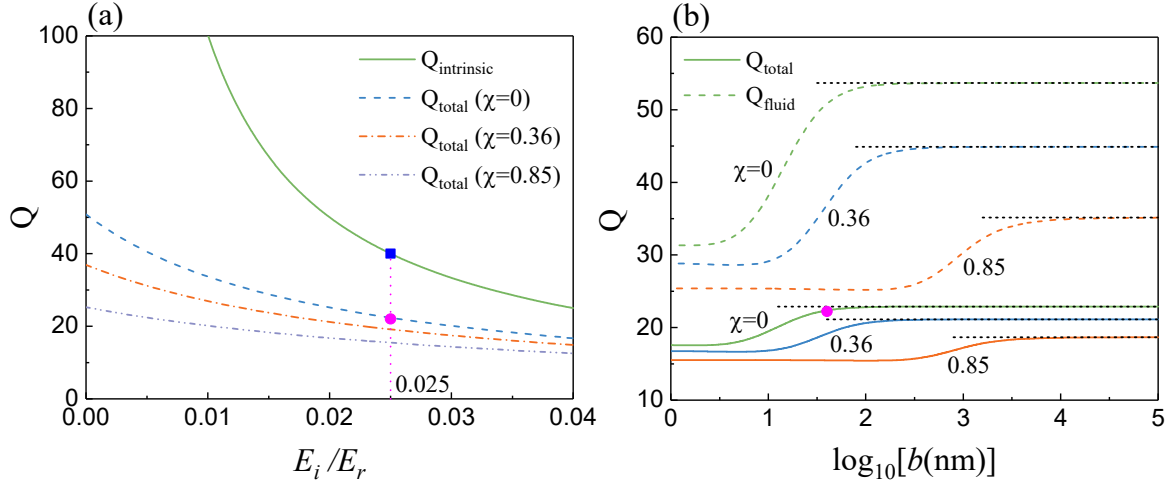

FIGURE 14. (a) Intrinsic quality factor $Q_{\text{intrinsic}}$ and total quality factor $Q_{\text{total}}$ of the breathing mode as functions of the loss-to-storage modulus ratio $E_i/E_r$ for a 40nm isotropic gold nanosphere vibrating in vacuum and in different glycerol-water mixtures. (b) Total quality factor $Q_{\text{total}}$ (solid lines) for $E_i/E_r = 0.025$ and fluid quality factor $Q_{\text{fluid}}$ (dashed lines) without solid intrinsic damping as functions of the logarithmic radius $\log_{10}[b(\text{nm})]$ for the breathing mode of isotropic gold nanospheres in different glycerol-water mixtures. In (a) and (b) the glycerol-water mixtures are described by the compressional non-Newtonian model, and the experimentally measured $Q_{\text{intrinsic}}$ (blue square) and $Q_{\text{total}}$ (pink circle) values conducted by Ruijgrok *et al.* (2012) are also displayed for the case of water ($\chi = 0$). In (b) the short-dashed lines represent the non-viscous compressible results.

First of all, for the breathing mode of a 40nm isotropic gold nanosphere vibrating in vacuum and glycerol-water mixtures, figure 14(a) shows the variations of $Q_{\text{intrinsic}}$ and $Q_{\text{total}}$ with the loss-to-storage modulus ratio $E_i/E_r$. As expected, both $Q_{\text{intrinsic}}$ and $Q_{\text{total}}$ decline monotonically with increasing $E_i/E_r$ owing to the enhancement of solid intrinsic damping. Note that Ruijgrok *et al.* (2012) carried out the experimental measurement of damping for the fundamental breathing mode of 40nm gold nanospheres optically trapped in water ($\chi = 0$). The measured average intrinsic and total quality factors are $Q_{\text{intrinsic}} = 40 \pm 4$ and $Q_{\text{total}} = 22 \pm 1.5$, respectively, both highlighted in figure 14(a). It is obvious that the theoretically predicted intrinsic and total quality factors agree exceptionally well with those measured experimentally (Ruijgrok *et al.* 2012). In addition, the loss-to-storage modulus ratio for gold nanospheres in experiment can be approximately determined as $E_i/E_r \simeq 0.025$ based on the present theoretical predictions. Interestingly, when $E_i/E_r$ increases, the variation range of $Q_{\text{total}}$ with the glycerol mass fraction gradually narrows, which is physically understandable that the solid intrinsic damping dominates in the quality factor.

Furthermore, the variations of $Q_{\text{total}}$ and $Q_{\text{fluid}}$ with the logarithmic radius $\log_{10}[b(\text{nm})]$ are depicted in figure 14(b) for the breathing mode of the isotropic gold nanosphere in different glycerol-water mixtures. Note that the loss-to-storage modulus ratio is taken as 0.025 in order to calculate $Q_{\text{total}}$, and the glycerol-water mixtures are described by the compressional non-Newtonian model. It can be found that the qualitative variation trend of $Q_{\text{total}}$ is essentially unaltered when incorporating the solid intrinsic damping into the FSI system. Thus, we can refer to the descriptions for the fluid quality factor $Q_{\text{fluid}}$ in Sec. 5.3 to explain similar phenomena for $Q_{\text{total}}$ and do not repeat here. Again, for a fixed glycerol mass fraction, the variation range of $Q_{\text{total}}$ with the nanosphere size is remarkably smaller than that of $Q_{\text{fluid}}$ because of the increasing solid intrinsic damping. For different glycerol-water mixtures and various vibration modes, table S2 in SM-V-E provides the theoretically predicted total and fluid quality factors for the reference of experimentalists.

## 6. Conclusions

In this paper, we analytically conducted the 3D free vibration analysis of a multilayered anisotropic sphere submerged in a compressible non-Newtonian fluid with both the shear and compressional relaxation processes. Firstly, we solved the linearized Navier-Stokes equation of the compressional non-Newtonian fluid for small-amplitude harmonic oscillations by employing three appropriate potential functions. Secondly, utilizing the state-space method which combines the state-space formalism and approximate laminate technique, we established two global transfer relations which correlate the state vectors at the outermost surface with those at the innermost surface of the anisotropic multilayered hollow sphere. Thirdly, we obtained two separate analytical frequency equations characterizing two independent classes of vibration (spheroidal and torsional modes) by incorporating the fluid-solid interface continuity conditions. Finally, based on the derived characteristic equations and the Muller's method to searching for complex roots, we provided a detailed numerical evaluation to comprehensively elucidate the effects of fluid viscosity and compressibility, fluid viscoelasticity, solid anisotropy and surface effect, as well as solid intrinsic damping on the vibration characteristics of the FSI system.

From the numerical results, we obtained the following important observations:

1) The proposed state-space method combined with the Muller's method can be used to obtain accurate dynamic responses of the FSI system with a high precision.

2) The vibration behavior of the sphere exhibits size-dependent characteristics owing to the presence of an intrinsic size for viscous fluids even if there is no surface effect. The attenuation behavior of a macro- or nano-sphere vibrating in a Newtonian fluid mainly depends on the fluid compressibility or viscosity.

3) For complex fluids with strong viscoelastic effect and for the compression-dominated flow, the compressional relaxation effect is extremely vital to correctly predict the FSI vibration characteristics at nanoscale.

4) For the nanoparticle of a given size, a critical fluid mass fraction exists, above which the vibration quality factor remains almost unchanged or even increases reversely due to the enhanced elastic effect in the fluid. Similarly, there exists a critical size of nanoparticle for a prescribed non-Newtonian fluid, below which the fluid elastic effect on the FSI vibration behavior becomes significant.

5) For higher-order modes or smaller spherical radii, for which high vibration frequencies are generally encountered, the viscoelastic effect on fluid-solid coupled vibrations becomes significant even in low-viscosity fluids such as pure water.

6) The anisotropy of the gold nanosphere does not qualitatively change the variation trend of the FSI vibration characteristics owing to its weak anisotropy. Even so, the deviation of the vibration frequency or quality factor for some particular modes (e.g. the dipolar mode $n=1$) may become significant and cannot be overlooked in practical applications.

7) A stiffer surface layer of nanospheres enhances the overall stiffness of the FSI system, makes more energy be stored in the nanosphere, and hence raises the vibration frequency and quality factor. A heavier surface layer of nanospheres has a tendency to weaken the fluid loading effect and thus heightens the quality factor.

8) The proposed method can accurately predict the quality factor of the submerged nanoparticles with intrinsic damping, and the results are consistent with the available experiments.

All these results lay a theoretical foundation for instances for 1) exploiting the FSI vibrations on the nanoscale to measure the viscoelastic properties of low-viscosity fluids with faster relaxation times, 2) utilizing the surface-dependent FSI vibrations at nanoscale to design new molecular and atomic sensors with high sensitivity, 3) actively tuning the FSI vibration characteristics by surface engineering, and 4) exploring the most appropriate vibration modes to destroy viruses. Indeed, these aspects are worthy of being further investigated by means of numerical simulations and experiments.

Finally, it should be emphasized here that the state-space method can be conveniently applied to other geometrical configurations such as layered nanoplates (Chang *et al.* 2015) and nanocylinders with or without interactions with complex fluids. In addition, wave propagation and vibration characteristics of nanostructures with multifield coupling effects (such as piezoelectric effect) working in complex fluids can be analyzed similarly using the proposed state-space method.

**Acknowledgements**

This work was supported by the National Natural Science Foundation of China (Nos. 11872329, 11621062, and 11532001). Partial supports from the Fundamental Research Funds for the Central Universities (No. 2016XZZX001-05) and the Shenzhen Scientific and Technological Fund for R & D (No. JCYJ20170816172316775) are also acknowledged.